\documentclass[floatfix,showpacs,superscriptaddress,prl,twocolumn]{revtex4}
\usepackage{amsbsy,amssymb,amsmath,bm}
\usepackage{epsf}
\usepackage{graphicx,color}
\usepackage{amsfonts}
\usepackage{amssymb}
\usepackage{amsmath}

\input{tcilatex}
\begin{document}

\title{Nonlocal electrodynamics in Weyl semi-metals}
\author{B. Rosenstein}
\affiliation{\textit{Electrophysics Department, National Chiao Tung University, Hsinchu
30050,} \textit{Taiwan, R. O. C}.}
\affiliation{\textit{Physics Department, Ariel University, Ariel 40700, Israel}}
\author{H.C. Kao}
\affiliation{\textit{Physics Department, National Taiwan Normal University, Taipei 11677,
Taiwan, R. O. C}.}
\author{M. Lewkowicz}
\affiliation{\textit{Physics Department, Ariel University, Ariel 40700, Israel}}
\email{vortexbar@yahoo.com}
\date{\today }

\begin{abstract}
Recently synthesized 3D materials with Dirac spectrum exhibit peculiar
electric transport qualitatively different from its 2D analogue, graphene.
Neglecting impuritiy scattering, the real part of the conductivity is
strongly frequency dependent (linear), while the imaginary part is non-zero
(unlike in undoped, clean graphene). The Coulomb interaction between
electrons is unscreened as in a dielectric and hence is long range. We
demonstrate that the interaction correction renders the electrodynamics
nonlocal on a mesoscopic\ scale. The longitudinal conductivity $\sigma _{L}$
(related by charge conservation to the electric susceptibility) and the
transverse conductivity $\sigma _{T}$ are different in the long wave length
limit and consequently the standard local Ohm's law description does not
apply. This leads to several remarkable effects in transport and optical
response. \ We predict a charging effect in DC transport that is a direct
signature of the nonlocality. The optical response of the WSM is also
sensitive to the nonlocality. In these materials p-polarized light generates
bulk plasmons as well as the transversal waves. The propagation inside the
WSM is only slightly attenuated. At a specific (material parameter
dependent) frequency the two modes coincide, a phenomenon impossible in a
local medium. Remarkably, for any frequency there is an incident angle where
total absorption occurs, turning the WSM opaque.
\end{abstract}

\maketitle

One of the common assumptions of electrodynamics in electrically active
media is that the effect of external electric fields can be described 
\textit{locally} by constitutive relations connecting the "induced" currents
to the electric field even when spatial dispersion is present. Generally,
due to space - time translational symmetry of the material, the relation
between Fourier components ($\omega $ is the frequency, $\mathbf{k}$ the
wavevector) of the electric field and these of the induced current density
within linear response reads:

\begin{equation}
J_{i}\left( \omega ,\mathbf{k}\right) =\sigma _{ij}\left( \omega ,\mathbf{k}%
\right) E_{j}\left( \omega ,\mathbf{k}\right) \text{.}  \label{induced}
\end{equation}%
Here $\sigma $ is the AC conductivity tensor with indices $i,j=x,y,z$. The
locality of the electrodynamic response in Fourier space means that the long
wavelength limit exists: $\sigma _{ij}\left( \omega ,\mathbf{k=0}\right)
\equiv \sigma _{ij}\left( \omega \right) $. The conductivity tensor for a
homogeneous, isotropic, space and time-reversal invariant (nongyrotropic)
material at $\mathbf{k=0}$ simplifies into the simple form of Ohm's law: 
\begin{equation}
\sigma _{ij}\left( \omega \right) =\delta _{ij}\sigma \left( \omega \right) ,
\label{Ohm}
\end{equation}%
or $\mathbf{J}\left( \omega \right) \mathbf{=}\sigma \left( \omega \right) 
\mathbf{E}\left( \omega \right) $.

On the microscopic level the locality is not guaranteed\cite{Vignale}. It
hinges on the nature of the charges in the condensed matter system and
presence of long-range interaction between them. These in turn determine the
long-wave excitations of the material. In an insulator (or semiconductor at
low temperatures) locality is simply a result of absence of gapless charged
excitations. This does not apply to metals.

In the free electron gas model of an ideal metal, i.e. neglecting both
disorder and electron-electron interactions, there is no energy gap, so
there are gapless charged excitations. The conductivity tensor in a metal
can be uniquely decomposed into a transversal and a longitudinal part
(assuming rotational and reflection symmetry for simplicity)\cite{Vignale}: 
\begin{equation}
\sigma _{ij}\left( \omega ,\mathbf{k}\right) =\left( \delta _{ij}-\frac{%
k_{i}k_{j}}{k^{2}}\right) \sigma _{T}\left( \omega ,\mathbf{k}\right) +\frac{%
k_{i}k_{j}}{k^{2}}\sigma _{L}\left( \omega ,\mathbf{k}\right) \text{,}
\label{decomposition}
\end{equation}%
Yet, the standard 'Lindhard' type calculations in an absolutely clean metal
with an arbitrary dispersion relation of the charge carriers and \textit{%
finite area} Fermi surface, see Fig.1a, show that the two scalars $\sigma
_{T}$ and $\sigma _{L}$ are not independent at small wavevectors, 
\begin{eqnarray}
\sigma _{T}\left( \omega ,\mathbf{k=0}\right) &=&\sigma _{L}\left( \omega ,%
\mathbf{k=0}\right) \equiv \sigma \left( \omega \right) ;  \label{betadif} \\
\sigma _{L}\left( \omega ,\mathbf{k}\right) -\sigma _{T}\left( \omega ,%
\mathbf{k}\right) &=&\beta \left( \omega \right) k^{2}+O\left( k^{4}\right) 
\text{,}  \notag
\end{eqnarray}%
thus leading to the local Ohm's law Eq.(\ref{Ohm}).

Impurities in a metal also define a length scale, the mean free path, that
effectively makes the carriers' motion diffusive and the charge excitations
"massive" in terms of their dispersion relation. Accounting for impurities
within the self consistent harmonic approximation (yet still neglecting
Coulomb interactions) thus makes the long wavelength limit of the Lindhard
diagram smooth\cite{Rammer}. Beyond the harmonic approximation, composite
excitations like diffusons have zero modes. However coupling of the external
fields to these excitations is "soft" enough to cause the so-called
"infrared divergencies" that in principle could make the long wavelength
limit singular \cite{Rammer}.

Returning to a very clean electron gas, the nonlocality can in principle
arise due to long-range Coulomb interactions. However, long-range
interactions are not screened only in insulators. In a metallic state with
finite density of charges (finite density of states on the Fermi surface)
the nonlocality is prevented by the screening of the Coulomb force that
becomes effectively short-range and thus unable to cause infrared
divergencies. As a result clean interacting charged electron gas is local.

All the above reasons ensuring locality, namely a direct energy gap,
significant disorder and screening of the Coulomb interactions, are
inapplicable to the recently discovered Weyl semi-metals (WSM). In these
crystals electronic states are described by the Bloch wave functions,
obeying the 3D "pseudo-relativistic" Weyl (Dirac) equation with Fermi
velocity $v$ replacing the velocity of light, see Fig.1b. The novelty of the
physics of WSM at Dirac point is clearly due to its "ultrarelativistic"
dispersion, $\varepsilon _{\mathbf{k}}=vk$, of the elementary excitations.
The Fermi level at neutrality point is occupied by a finite number of states
(zero density of states), so that the excitations are those of the neutral
plasma.

Although the two-band electronic structure of bismuth was described by a
four-component nearly massless Dirac fermion in 3D caused by spin-orbit
interaction long ago \cite{Wolff} (with spin replacing pseudospin), only
recently several systems were experimentally demonstrated to exhibit the 3D
Dirac quasiparticles. Materials are quite diverse and include the time -
reversal invariant topological Dirac semimetal \cite{Liu1,Kushwaha} $%
Na_{3}Bi $ (predicted\cite{Wang13a,Wang13b}), a bulk crystal symmetry
protected semimetal\cite{Fang,Liu2} $Cd_{3}As_{2}$ with a single pair of
Dirac points \cite{Neupone}, crystals on the phase transition boundary
between topological and band insulators $HgCdTe$\cite{Potemski}. Recently,
Weyl semi-metals exhibiting Fermi arcs on their surface were discovered\cite%
{Lv1,Lv2,Xu}. All these materials exhibit a great variety of new
electromagnetic transport and optical phenomena (not seen even in 2D WSM
like graphene) including giant diamagnetism, quantum magnetoresistance
showing linear field dependence\cite{Ogata,Ogata12,Xiong,Liang},
superconductivity\cite{Wang15} \textit{etc.}

There is no finite energy gap in WSM to ensure locality as in band
insulators or semiconductors. In this paper we assume that the WSM is clean
enough, so that the disorder scale is irrelevant and the chemical potential
is tuned to the Dirac point. The elementary excitations are still "massless"
at long wavelengths like in a metal. To no surprise the free
(noninteracting) electron gas AC conductivity of WSM, calculated recently%
\cite{RosensteinPRB13,Zhang13}, turned out to remain local (see discussion
below). However calculating the screening due to the interaction corrections%
\cite{Vishwanath,RosensteinPRB13,DasSarma-arxiv}, it became clear that the
Coulomb interactions are still long range (namely like an insulator WSM is
unable to screen the Coulomb interactions, unlike in the metal). The reason
for locality mentioned with respect to metals therefore does not apply. So
that there is no good reason to exclude the possibility that the long-wave
electric response of WSM in the presence of strong interactions is nonlocal.
To summarize WSM exhibits a curious mixture of properties usually attributed
either to metals or insulators that does not allow to apply a conventional
reasoning to establish locality of the linear response.

In the present paper we demonstrate by explicit microscopic calculation of
the interaction effects in a rather generic model of WSM that the
electrodynamics is indeed nonlocal. The rather unusual macroscopic
electrodynamics then is formulated and applied to various physical
phenomena. Several experimental setups in which the nonlocality can be
demonstrated are suggested. These include the charging effects in DC
transport, while another is the optical generation of both longitudinal
(plasmon) and transversal waves in these materials and their subsequent
propagation.

\section{Results}

\subsection{The model}
An analogous calculation in graphene\cite{Rosenstein13,RosensteinPRB14} (a
2D version of WSM) reveals that in order to avoid complications linked to
the absence of scale separation in the system of relativistic massless
fermions (known otherwise as "anomaly"), one should use a well defined
lattice model. Electrons in WSM are described sufficiently accurately for
our purposes by the tight binding model of nearest neighbors on a cubic
lattice\cite{RosensteinPRB13, Mastropietro} $\mathbf{n}=n_{i}\mathbf{a}_{i}$
(see Fig.2a). The Hamiltonian is 
\begin{equation}
\widehat{K}=\frac{i\gamma }{2}\dsum\limits_{n,i}c_{\mathbf{n}}^{\alpha
\dagger }\sigma _{i}^{\alpha \beta }c_{\mathbf{n}+\mathbf{a}_{i}}^{\beta }+hc%
\text{,}  \label{K}
\end{equation}%
where $\sigma _{i}$ are Pauli matrices, operators $c_{\mathbf{n}}^{\alpha
\dagger }$ , $\alpha =1,2$, create a two\ - component spinor (describing two
orbitals per site) and $\gamma $ is the hopping energy determining the Fermi
velocity $v=\gamma a/\hbar $ ($a$ - lattice spacing). This defines a
propagator depicted by arrows (see Fig.2b) in the Feynman diagrams. The
Coulomb interaction (neglecting retardation effects) on the lattice can be
viewed as an exchange of static photons. In term of Fourier components it
takes the form (see\cite{Sup1} for details) $\ $

\begin{equation}
\widehat{V}=\frac{1}{2\mathcal{V}}\sum\limits_{\mathbf{pkl}}v_{\mathbf{p}}c_{%
\mathbf{k+p}}^{\sigma \dagger }c_{\mathbf{k}}^{\sigma }c_{\mathbf{l-p}%
}^{\rho \dagger }c_{\mathbf{l}}^{\rho }\text{,}  \label{V}
\end{equation}%
where $\mathcal{V}$ is the sample volume and 
\begin{equation}
v_{\mathbf{p}}=\frac{\pi e^{2}}{\sin ^{2}\left( p_{x}a/2\right) +\sin
^{2}\left( p_{y}a/2\right) +\sin ^{2}\left( p_{z}a/2\right) }\text{.}
\label{v}
\end{equation}%
The photon propagator $v_{\mathbf{p}}$ is represented by wavy lines in Fig.2.

It is convenient to calculate the transverse and longitudinal conductivities
from the scalar quantities using $\sigma _{L}=k_{i}k_{j}\sigma _{ij}/k^{2}$
and $\sigma _{ii}=2\sigma _{T}+\sigma _{L}$. The relations follow from Eq.(%
\ref{decomposition}). The calculation involves evaluation of diagrams 2c for
free electrons and 2d, 2e, 2f for the leading interaction corrections, self
energy, the vertex renormalization and "glasses" diagram, respectively.
Details involving renormalization, cancellation of both infrared and
ultraviolet divergences appear in the Supplementary Information SI1, while
results are given Eq.(\ref{sig0}) and Eq.(\ref{signl}).

\subsection{The electrodynamics of a free Fermi gas is local}

The dispersive conductivity tensor for a free (neglecting the
electron-electron interaction) clean ultra-relativistic fermion gas at Dirac
point is (see \cite{Sup1}):%
\begin{widetext}
\begin{equation}
\sigma _{ij}\left( \omega ,\mathbf{k}\right) =\frac{Ne^{2}}{24\pi \hbar
v\omega }\left( 1-\frac{i}{\pi }\log \frac{\Lambda ^{2}v^{2}}{\omega
^{2}-v^{2}k^{2}}\right) \left\{ \delta _{ij}\left( \omega
^{2}-v^{2}k^{2}\right) +v^{2}k_{i}k_{j}\right\} \text{.}  \label{sig0}
\end{equation}%
\end{widetext}%
$N$ is the number of Weyl fermions and $\Lambda \sim 1/a$ is the ultraviolet
cutoff. In the long wavelength limit one recovers the AC conductivity $%
\sigma _{ij}^{0}\left( \omega ,\mathbf{k=0}\right) =\sigma _{0}\left( \omega
\right) \delta _{ij}$ with%
\begin{equation}
\sigma _{0}\left( \omega \right) =\frac{Ne^{2}\omega }{24\pi \hbar v}\left(
1-\frac{i}{\pi }\log \frac{\Lambda ^{2}v^{2}}{\omega ^{2}}\right) \text{.}
\label{sigma0}
\end{equation}%
This has both a real and an imaginary part, logarithmically divergent as
function of $\Lambda $. Thus the DC conductivity is zero, i.e. the material
behaves like an insulator, qualitatively different from graphene that is a
pseudo-dissipative metal. This already indicates that in 3D the Coulomb
interaction is unscreened\cite{RosensteinPRB13}. The dependence on the wave
vector follows uniquely from the pseudo- relativistic invariance of the free
Weyl gas like in its 2D analogue, graphene \cite{RosensteinPRB14}. Hence,
using notations of Eq.(\ref{betadif}), the electrodynamics is local with 
\begin{equation}
\beta ^{0}\left( \omega \right) =-\frac{\sigma _{0}\left( \omega \right)
v^{2}}{\omega ^{2}}\text{.}  \label{sig0L}
\end{equation}

\subsection{Interactions cause nonlocal electrodynamics}

Using the Coulomb interaction Hamiltonian within the tight binding model,
Eq.(\ref{V}), one obtains the corrections to first order in the effective
interaction strength $\alpha =e^{2}/\kappa \hbar v$, where $\kappa $ denotes
the dielectric constant of the background. While the corrections to either $%
\sigma _{T}$ or $\sigma _{L}$ are small, the relative correction to the
difference $\sigma _{nl}\equiv \sigma _{L}-\sigma _{T}$ is actually dominant
for $\mathbf{k=0}$ and can be considerable in the homogenous regime $%
v^{2}k^{2}<<\omega ^{2}$. While in free WSM $\sigma _{nl}\left( \omega ,%
\mathbf{0}\right) $ was zero, now it becomes finite: 
\begin{equation}
\sigma _{nl}\left( \omega ,\mathbf{0}\right) =\sigma _{0}\left( \omega
\right) \left\{ \frac{\pi \alpha }{3}\log \frac{0.6\omega ^{2}}{\overline{%
\omega }^{2}}+O\left( \alpha ^{2}\right) \right\} .  \label{signl}
\end{equation}%
Here $\overline{\omega }$ is the frequency at which the renormalized value
of the (renormalization group "running") Fermi velocity $v\left( \overline{%
\omega }\right) $ and the renormalized coupling $\alpha =\alpha \left( 
\overline{\omega }\right) $ are defined\cite%
{Vishwanath,RosensteinPRB13,DasSarma-arxiv}. Details of the calculation are
given in \cite{Sup1}.

\subsection{Optics\textbf{/}Plasmonics within WSM}

Let us now develop the equations for the electromagnetic fields on the
macroscopic scale in a WSM. We employ the Landau-Lifshitz\cite{LL8, Golubkov}
definition of the macroscopic fields

\begin{eqnarray}
\nabla \cdot \mathbf{D} &=&4\pi \rho _{ext};\text{ \ \ }\nabla \times 
\mathbf{E=-}\frac{1}{c}\mathbf{\dot{B};}  \label{Maxwell} \\
\nabla \cdot \mathbf{B} &=&0;\text{ \ \ }\nabla \times \mathbf{B}=\frac{1}{c}%
\left( \mathbf{\dot{D}}+4\pi \mathbf{J}_{ext}\right) \text{.}  \notag
\end{eqnarray}%
The displacement field is $\mathbf{D}=\mathbf{E}+4\pi \mathbf{P},$ where the
polarization vector is defined by $\mathbf{\dot{P}}=\mathbf{J}$. The induced
current Eq.(\ref{induced}) is determined nonlocally by the conductivity
tensor given in Eqs.(\ref{sig0L}) and (\ref{signl}). In particular, for the
harmonic dependence $\mathbf{E}\left( \mathbf{r},t\right) =\mathbf{E}%
_{0}e^{i\left( \mathbf{k\cdot r}-\omega t\right) }$ (and without external
charges and currents) Gauss' Law demands that either $\mathbf{E}$ is
transversal, $\mathbf{k}\cdot \mathbf{E=}0$, $\ $or else the dispersion
relation is plasmonic: 
\begin{equation}
1+i\frac{4\pi }{\omega }\sigma _{L}\left( \omega ,\mathbf{k}\right) =0.
\label{epsl}
\end{equation}%
Amp\`{e}re's Law together with Faraday's Law results in the condition that
either $\mathbf{E}$ is longitudinal, $\mathbf{k\times E=0,}$ or the
dispersion relation is transversal: 
\begin{equation}
1+i\frac{4\pi }{\omega }\sigma _{T}\left( \omega ,\mathbf{k}\right) =\frac{%
c^{2}k^{2}}{\omega ^{2}}\text{.}  \label{epst}
\end{equation}

It is interesting to note that when $\omega <<\Lambda v$ the conductivity,
Eq.(\ref{sig0L}) is approximately purely imaginary and one obtains the
dispersion relations for the longitudinal ("plasmon", wavevector $\mathbf{q}$%
), and transversal ("light", wavevector $\mathbf{p}$), waves :

\begin{eqnarray}
q^{2} &=&\frac{\omega ^{2}}{v^{2}}\left\{ 1+\left( 1+\frac{N\alpha }{24\pi }%
\log \frac{\Lambda ^{2}v^{2}}{\omega ^{2}}\right) ^{-1}+\frac{\pi \alpha }{3}%
\log \frac{0.6\omega ^{2}}{\overline{\omega }^{2}}\right\} \text{;}  \notag
\\
&&  \label{dispL} \\
p^{2} &=&\frac{\omega ^{2}}{c^{2}}\left\{ 1+\frac{1+\frac{N\alpha }{24\pi }%
\log \left( \Lambda ^{2}v^{2}/\omega ^{2}\right) }{1+\frac{v^{2}}{c^{2}}%
\frac{N\alpha }{24\pi }\log \left( \Lambda ^{2}v^{2}/\omega ^{2}\right) }%
\right\} \text{.}  \label{dispT}
\end{eqnarray}%
The results for $q,p$ in units of $\overline{\omega }/v$ as functions of $%
\omega /\overline{\omega }$ in the case of $N=4$, $v\left( \overline{\omega }%
\right) =c/300$, $\alpha \left( \overline{\omega }\right) =0.2$, $\hbar
v\Lambda =5eV$, (characteristic of \ $Na_{3}Bi$ and $Cd_{3}As_{2}$ for $%
\overline{\omega }$ in the $THz$ range)\ are presented in Fig. 3. One
observes that the light wavevector $\mathbf{p}$ is real for all frequencies,
that is the wave is nondissipating, see the left inset in Fig.3a. The
plasmon wave vector $\mathbf{q}$ is real, according to Eq.(\ref{dispL}),
only above a certain threshold frequency $\omega _{p}$, see Fig.3b. One
should emphasize that the "plasma frequency" is that of a neutral plasma
rather than that of the charged plasma in metals. However within the
renormalized perturbation theory used to derive the equation, the low
frequency region might be unreliable. Above the threshold the two waves
coexist and both wavevectors are real.

In local materials equations Eq.(\ref{dispT}) and Eq.(\ref{dispL}) cannot be
satisfied by the same wave vector: the difference of the equations is $i%
\frac{4\pi }{\omega }\sigma _{nl}\left( \omega ,\mathbf{q}_{m}\right)
=-c^{2}k^{2}/\omega ^{2}$. Using $\sigma _{nl}\left( \omega ,\mathbf{k}%
\right) =-\beta _{0}\left( \omega \right) k^{2},$see Eq.(\ref{sig0L}), this
leads to $4\pi i\omega \beta _{0}\left( \omega \right) =c^{2}$ that
generally cannot\ be satisfied. In WSM, on the other hand, $\sigma
_{nl}\left( \omega ,\mathbf{0}\right) $ is finite and determines the special
value of the wave vector at which both transverse and longitudinal waves are
the same. We further discuss this in the optical set up. Next we suggest \
two potential experimental setups that demonstrate the nonlocality of the
WSM electrodynamics.

\subsection{Reflection and refraction on a dielectric - WSM interface}

Let's consider the following setup, see Fig.4. An electromagnetic wave with
the wavevector $k_{x}>0,k_{y}=0$ and $k_{z}<0$ is incident at the angle $%
\theta $ with the normal on a semi-infinite WSM. The electric field with p -
polarization in vacuum is then described by a superposition of incoming and
reflected waves:%
\begin{widetext}
\begin{equation}
E_{i}^{vac}=A\left\{ \frac{k_{z}}{k},0,\frac{-k_{x}}{k}\right\} e^{i\left(
k_{x}x+k_{z}z\right) }-B\left\{ \frac{k_{z}}{k},0,\frac{k_{x}}{k}\right\}
e^{i\left( k_{x}x-k_{z}z\right) }  \label{Evac}
\end{equation}%
where $A$ is the incoming and $B$ the reflected amplitude, respectively. The
field in the WSM is given by%
\begin{equation}
E_{i}=C_{T}\left\{ \frac{p_{z}}{p},0,\frac{-k_{x}}{p}\right\} e^{i\left(
k_{x}x+p_{z}z\right) }+C_{L}\left\{ \frac{k_{x}}{q},0,\frac{q_{z}}{q}%
\right\} e^{i\left( k_{x}x+q_{z}z\right) },  \label{Ewsm}
\end{equation}%
\end{widetext}
where $C_{T}$ and $C_{L}$ are the transmitted transversal and longitudinal
amplitudes.

Employing the Landau-Lifshitz notation for the macroscopic fields in Maxwell
equations in dispersive media Eq.(\ref{Maxwell}), used by Golubkov\cite%
{Golubkov} to resolve similar problems as in exciton physics\cite{Agranovich}%
, the list of the continuity conditions on the boundary is:

(i) the normal component of $\mathbf{D}$:

\begin{equation*}
E_{z}^{vac}=\frac{4\pi i}{\omega }\sigma _{zx}E_{x}^{wsm}+\left( 1+\frac{%
4\pi i}{\omega }\sigma _{zz}\right) E_{z}^{wsm}.
\end{equation*}

(ii) the tangential component of $\mathbf{E}$: $E_{x}^{vac}=E_{x}^{wsm}$.

(iii) the normal component of $\mathbf{B}$\textbf{: }$\partial
_{x}E_{y}^{vac}=\partial _{x}E_{y}^{wsm}$.

(iv) the tangential component of $\mathbf{B}$: $\partial
_{z}E_{x}^{vac}-\partial _{x}E_{z}^{vac}=\partial _{z}E_{x}^{wsm}-\partial
_{x}E_{z}^{wsm}$.

(v) In addition, since two different modes propagate in the bulk one needs a
so-called ABC (additional boundary condition)\cite{Pekar, Agranovich}.
Assuming a "sharp" interface (width smaller than the wavelength) the
simplest ABC suffices:

\begin{equation*}
J_{z}\left( z=0\right) =0\rightarrow \sigma _{zx}E_{x}^{wsm}+\sigma
_{zz}E_{z}^{wsm}=0\text{.}
\end{equation*}%
This reflects the fact that the electric current cannot escape the material.

Applying these boundary conditions we obtain the following reflection and
transmission amplitudes:%
\begin{eqnarray}
r &=&\frac{1-D}{1+D};t_{T}=\frac{2k/p}{1+D};\text{ \ }t_{L}=-\frac{%
2k_{x}q\left( p^{2}-k^{2}\right) }{kq_{z}p^{2}\left( 1+D\right) }\text{ ;} 
\notag \\
D &=&\frac{k^{2}p_{z}q_{z}-\left( p^{2}-k^{2}\right) k_{x}^{2}}{%
k_{z}q_{z}p^{2}}\text{\ .}  \label{ppol}
\end{eqnarray}%
When $D=1$ the reflection coefficient vanishes and total absorption occurs,
thus turning the WSM opaque at frequency-dependent incident angles.

The s - polarization does not generate a plasmon. The amplitudes are
standard: $r=\left( 1-p_{z}/k_{z}\right) /\left( 1+p_{z}/k_{z}\right) $ and $%
t=2/\left( 1+p_{z}/k_{z}\right) $. In Fig.5a the amplitudes for the
p-polarizations (solid) and the s-polarizations (dashed) are presented.
Fig.5b shows the vanishing of the p-polarization reflection coefficient at
various incident angles in the $THz$ range.

\subsection{Surface charging due to DC current}

Two possible experiments are suggested for low frequency transport in WSM. A
Corbino geometry is employed in order to avoid ambiguities related to
boundaries, see Figs.6 and 7. A slowly alternating flux is generated by a
thin solenoid inserted into the cylindrical aperture (radius $R$) inside a
sufficiently large WSM sample. The fluxon's magnetic field is linear in time
(at times smaller than $1/\omega $). Within that period of time the
associated electric fields in the aperture and the WSM are time independent.
In the first setup the fluxon is concentrated at distance $d$ from the
center: $\Phi \left( \mathbf{r,}t\right) =\Phi _{0}\omega t\delta \left( 
\mathbf{r}-d\mathbf{\hat{x}}\right) $. The solenoidal part of the electric
field produced by the fluxon is 
\begin{equation}
E_{i}^{sol}\left( \mathbf{r}\right) =\frac{\omega \Phi _{0}}{2\pi b^{2}}%
\varepsilon _{ij}b_{j},  \label{Erot}
\end{equation}%
where $\mathbf{b}=\mathbf{r}-d\mathbf{\hat{x},}$ $i,j=x,y$. It induces
currents in the WSM that in turn generate charge distributions on the
aperture surface, provided that the flux is not concentric with the cylinder
axis. The surface charge density $Q\left( \phi \right) $ is obtained from
the condition that the normal component of the current on the surface on the
semimetal side vanishes, $\mathbf{n\cdot J=0}$. The irrotational part of the
electric field is

\begin{equation}
E_{i}^{irr}\left( \mathbf{r}\right) =L\int_{\phi =0}^{2\pi }\int_{z}\frac{%
s_{i}\text{ }Q\left( \phi \right) }{\left( z^{2}+s^{2}\right) ^{3/2}},
\label{Eirr}
\end{equation}%
where the distance from a point on the surface to the observation point $%
\mathbf{r}$ is denoted by $\mathbf{s}=\mathbf{r-}R\mathbf{n}^{\prime }$. The
boundary condition, $\mathbf{n\cdot J=n\cdot }\left( \sigma _{T}\mathbf{E}%
^{sol}+\sigma _{L}\mathbf{E}^{irr}\right) =0$, together with Gauss's Law, $%
\mathbf{n\cdot E}^{irr}\mathbf{=}2\pi Q\left( \phi \right) $, $\,$leads to a
dipole-like charge density. The details of the calculation are given in \cite%
{Sup2}. The surface charge density as a function of the polar angle $\phi $
is 
\begin{equation}
Q\left( \phi \right) =\frac{\sigma _{T}}{\sigma _{L}}\omega \Phi _{0}\frac{%
d\sin \phi }{4\pi ^{2}\left( R^{2}-2Rd\cos \phi +d^{2}\right) }\text{.}
\label{Q1}
\end{equation}%
The dominant linear $\omega $ dependence of the conductivities cancels in
the ratio $\frac{\sigma _{T}}{\sigma _{L}}=1-\frac{\pi \alpha }{3}\log \frac{%
0.6\omega ^{2}}{\overline{\omega }^{2}}$. The surface charge can be measured
by gauging the electric field near the interior surface of the material. The
deviation of $\sigma _{T}/\sigma _{L}$ from unity (the value in the case of
a local conductor) is an indicator of nonlocality of electrodynamics. Note
that a central position of the fluxon would not induce the surface charges.
The distribution of charge is given in Fig. 6b.

for surface charging due to DC current, the calculation is more complex
since induced charges appear not only on the aperture surface, but also on
the

For the second experimental setup, see Fig. 7 involve a Corbino disk with
the fluxon in the center of a composite cylinder of WSM $\left( 0<\phi <\pi
\right) $ and a usual "local" conductor (for example semiconductor) in the
lower segment $\pi <\phi <2\pi $; surface charges are now induced at the
interfaces $\phi =0$ and $\pi $ between the two materials which in turn
cause surface charges at the aperture. This results in an integral equation
that fortunately can be solved exactly, see \cite{Sup2}, and results in 
\begin{equation}
Q\left( \phi \right) =\frac{\sigma -\sigma _{T}}{\sigma +\sigma _{L}}\frac{%
\omega \Phi _{0}}{\left( 2\pi \right) ^{3}R}\log \frac{1+\cos \phi }{1-\cos
\phi }.  \label{Q2}
\end{equation}%
The distribution of this surface charge density is shown in Fig.7.

\section{Discussion}

The calculation of the leading Coulomb interaction effect on the
electromagnetic response of 3D Weyl semimetal reveals that its macroscopic
electrodynamics becomes nonlocal in a sense that the wave vector dependent
AC conductivity tensor becomes nonanalytic at small wave vectors, see Eq.(%
\ref{decomposition}). The origin of nonlocality is a unique combination of
the long - range (unscreened) Coulomb interactions like in dielectrics and
the ultra - relativistic nature of the quasiparticles. The longitudinal
conductivity $\sigma _{L}$ (related by charge conservation to the electric
susceptibility) and the transverse conductivity $\sigma _{T}$ are different
in the long wave length limit and consequently the standard local Ohm's law
description does not apply.

Several experimental signatures of the nonlocal contribution to
conductivity, Eq.(\ref{signl}) were pointed out. After reformulating the
macroscopic electrodynamics in terms of the longitudinal and transverse
fields with corresponding material parameters, the charging effect in DC
transport in Corbino geometry was worked out. The set up avoids potential
complications with conventional leads and contacts, while allowing for an
unambiguous indication of nonzero difference $\sigma _{nl}=\sigma
_{L}-\sigma _{T}$. A more conventional set up is the measurement of optical
response of the WSM that is also sensitive to the nonlocality. The
p-polarized light at general frequency $\omega $ generates in these
materials besides the transversal waves also the longitudinal one (that can
be viewed as a charge density wave or a bulk plasmon). The propagation
inside the WSM is only slightly attenuated, so that one can measure both the
reflection and two transmission amplitudes, all dependent on $\sigma _{nl}$.
At a specific (material parameter dependent) frequency the two modes
coincide, a phenomenon impossible in a local medium. The polarization
dependence of the light reflection (transmission) off a WSM slab is
sensitive to both $\sigma _{L}$ and $\sigma _{T}$. Remarkably for any
frequency there exists an incident angle where total absorption occurs, so
that WSM is opaque.

The applicability of perturbation theory in 3D Weyl semi-metals was quite
recently addressed by several groups who used renormalization group and
other nonperturbative methods like the random phase and large $N$
approximations\cite{Gonzales,Nagaosa,DasSarma-arxiv}. The latter shows
perturbation theory to be reliable over a wide range of values of the
interaction strength $\alpha $, up to a critical value $\alpha _{c}\approx
14.$ In addition, typical values of the background dielectric constant in 3D
WSM in the range $\kappa \sim 20-40$ have been measured\cite{DasSarma-arxiv}%
, which assures a below the critical value of the interaction strength. In
contrast, in the 2D case, where a critical value $\alpha _{c}=0.78$ was found%
\cite{DasSarmaPRB89},\ the situation for perturbation theory is less
fortunate since the measurement on samples substrated on boron nitride or on
suspended samples provided values close to that\cite{Siegel,Yu}.

\textit{Acknowledgements.} We are indebted to W.B. Jian, I. Herbut, E.
Farber, C. W. Luo, D. Cheskis. Work of H.K. was supported by NSC of R.O.C.
Grants101-2112-M-003-002-MY3.

\begin{figure*}
\vskip-1.0cm\hskip -12.0cm
\includegraphics[width=0.20\textwidth]{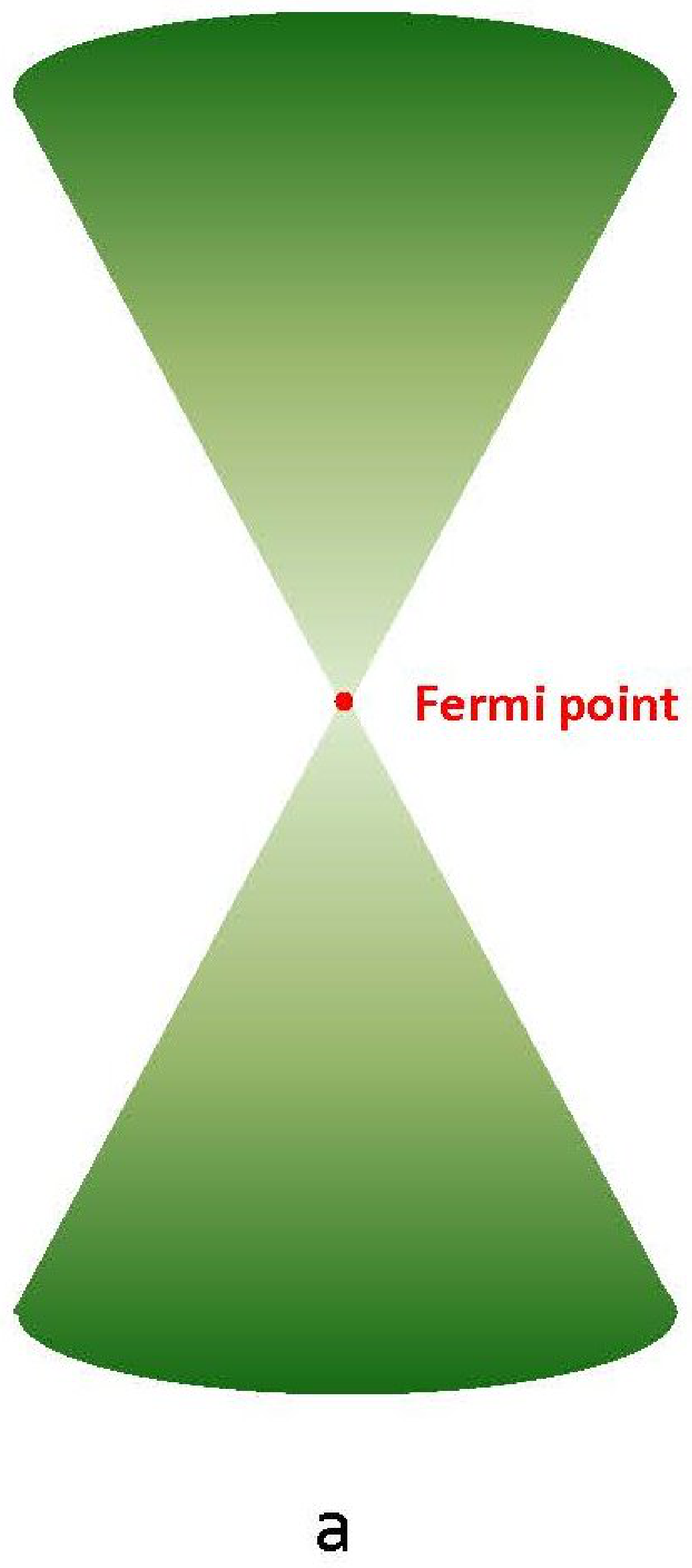}
\vskip -5.6cm\hskip 30.0cm
\includegraphics[width=0.25\textwidth]{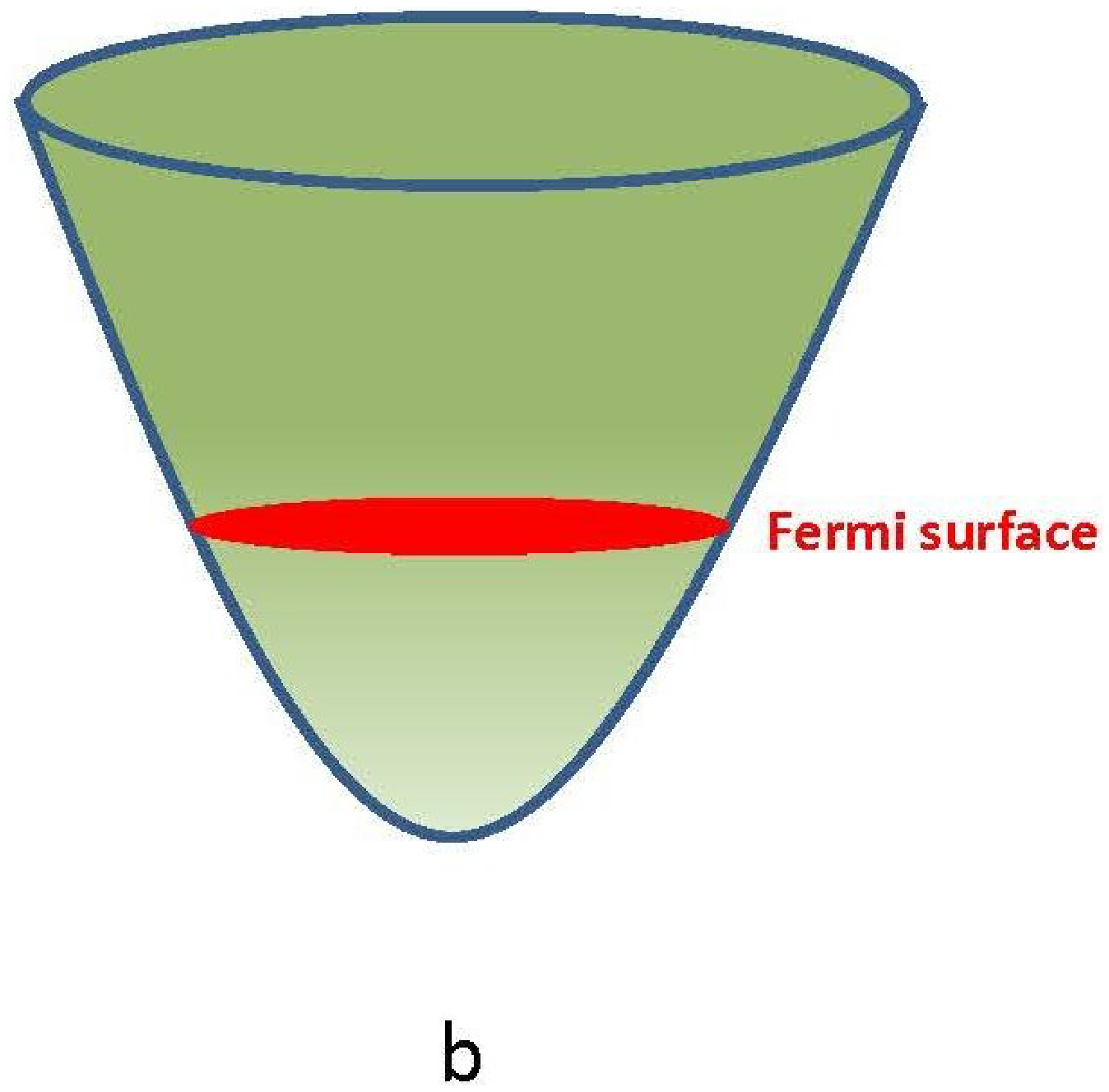}
\vskip 1.0cm
\caption{Spectrum of a charged metal contrasted to that of a Weyl semimetal.\hfil\break
a. The Fermi level surface area is nonzero for a metal (or a band
semimetal). The density of states at the Fermi level is finite (charged
plasma) as is the density of charge carriers.\hfil\break
b. In a Weyl semimetal at neutrality (Dirac) point the Fermi level crosses
the spectrum at finite number of states, so that the Fermi level surface
area is zero. The electron gas constitutes a neutral plasma.\hfil\break}
\end{figure*}

\begin{figure*}
\vskip-0.5cm\hspace{-12.0cm} 
\includegraphics[width=0.20\textwidth]{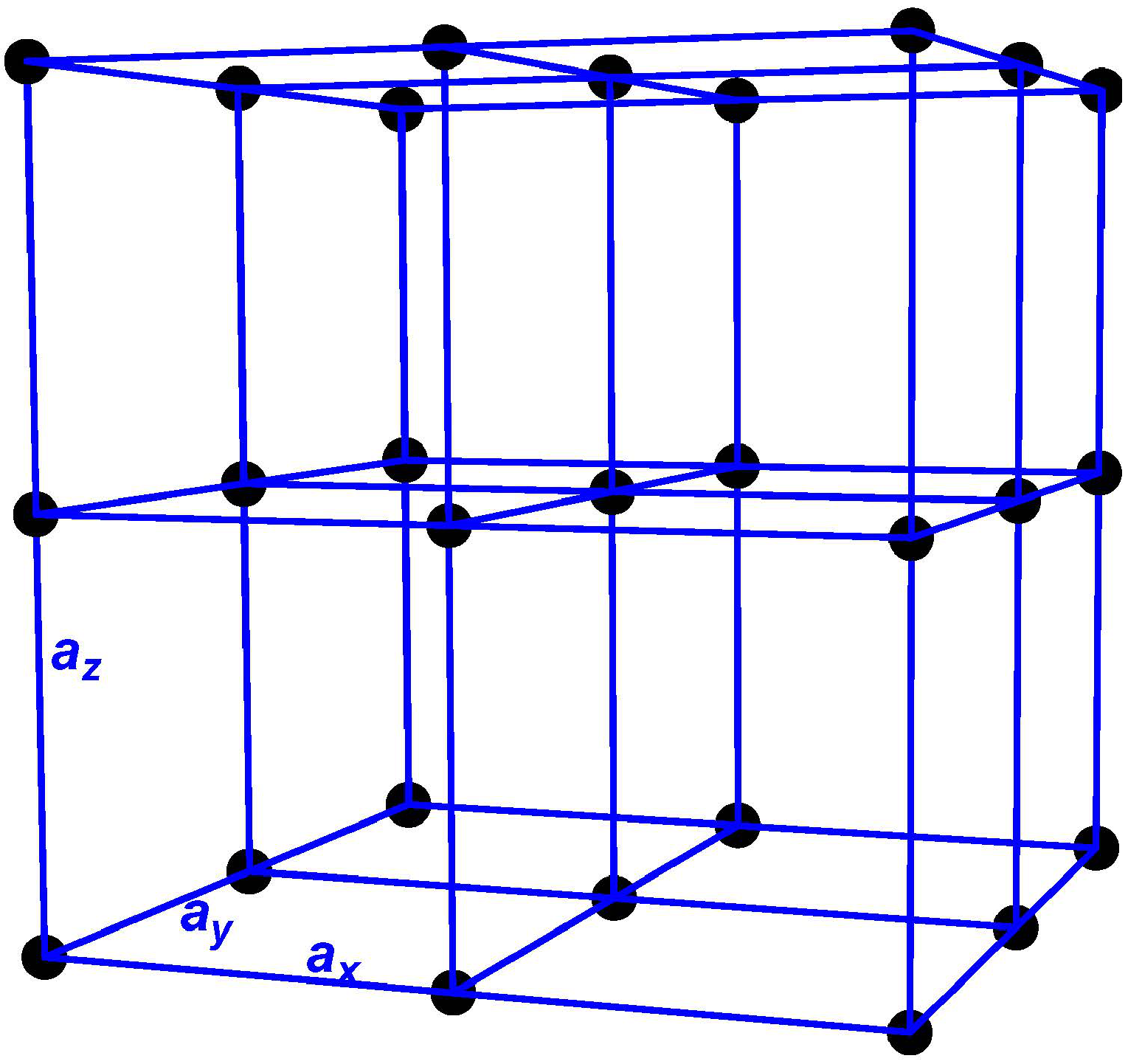}
\vskip-3.6cm\hspace{15.0cm}
\includegraphics[width=0.20\textwidth]{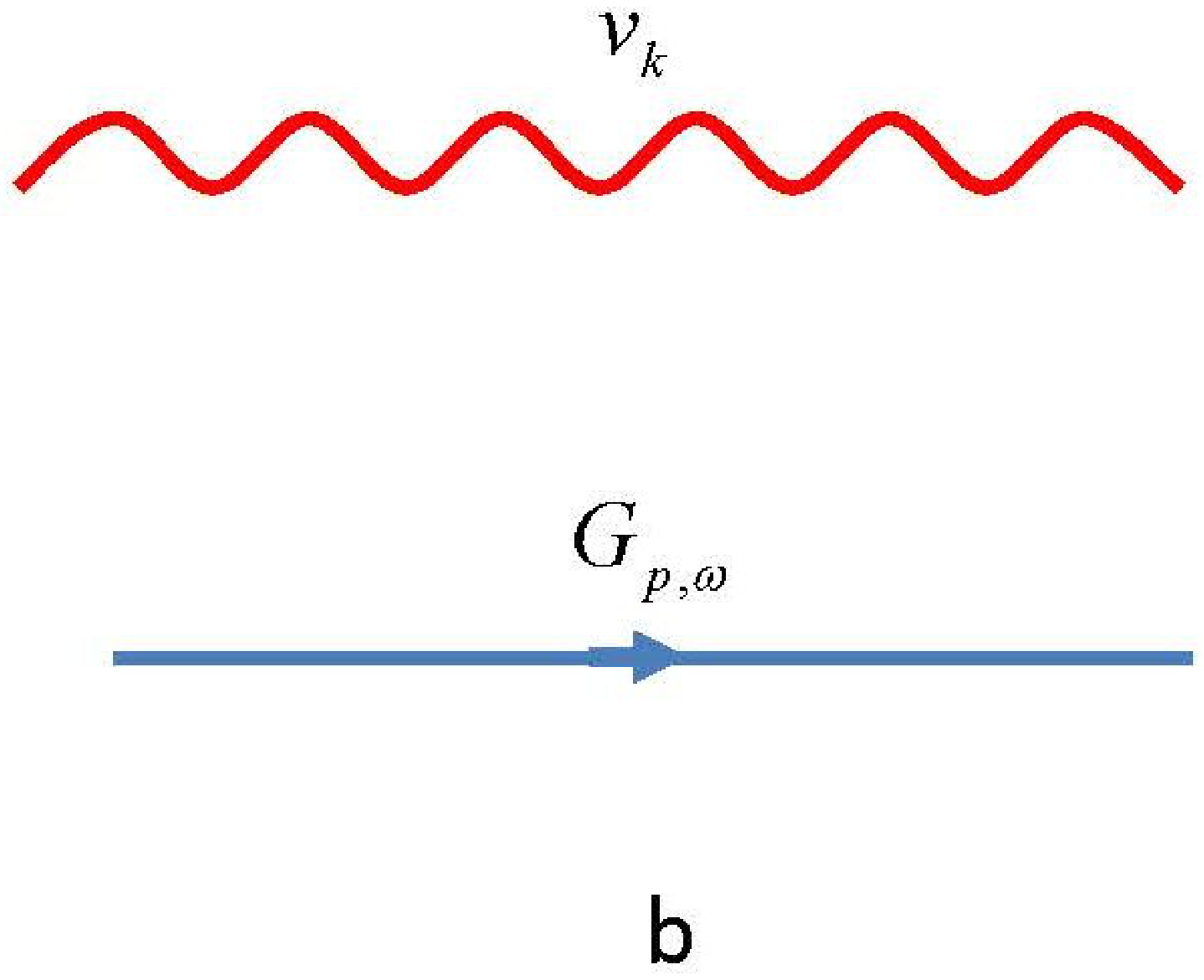}
\vskip0.6cm\hspace{-12.0cm}
\includegraphics[width=0.20\textwidth]{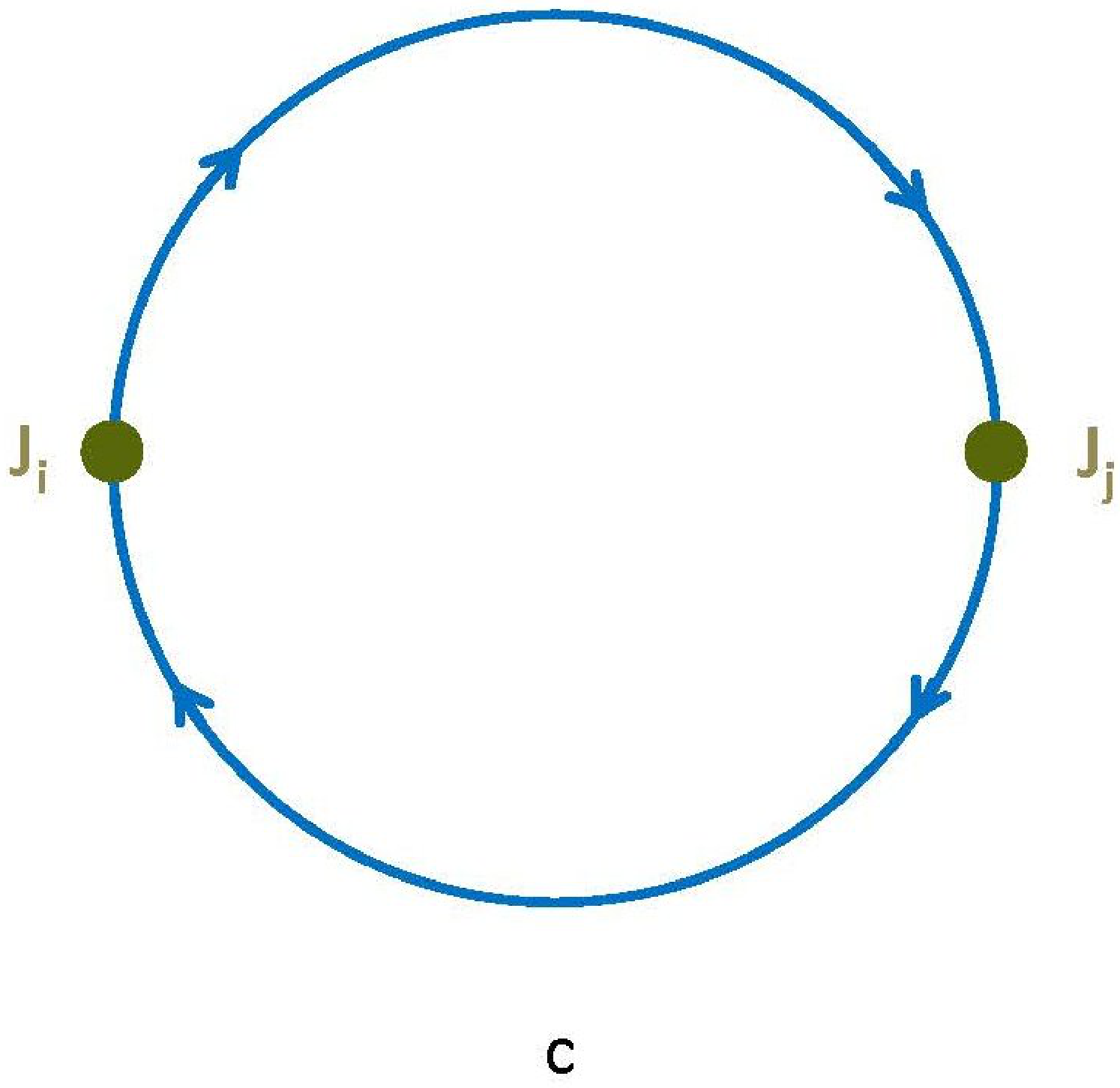}
 \vskip-4.0cm\hskip 15.0cm
\includegraphics[width=0.20\textwidth]{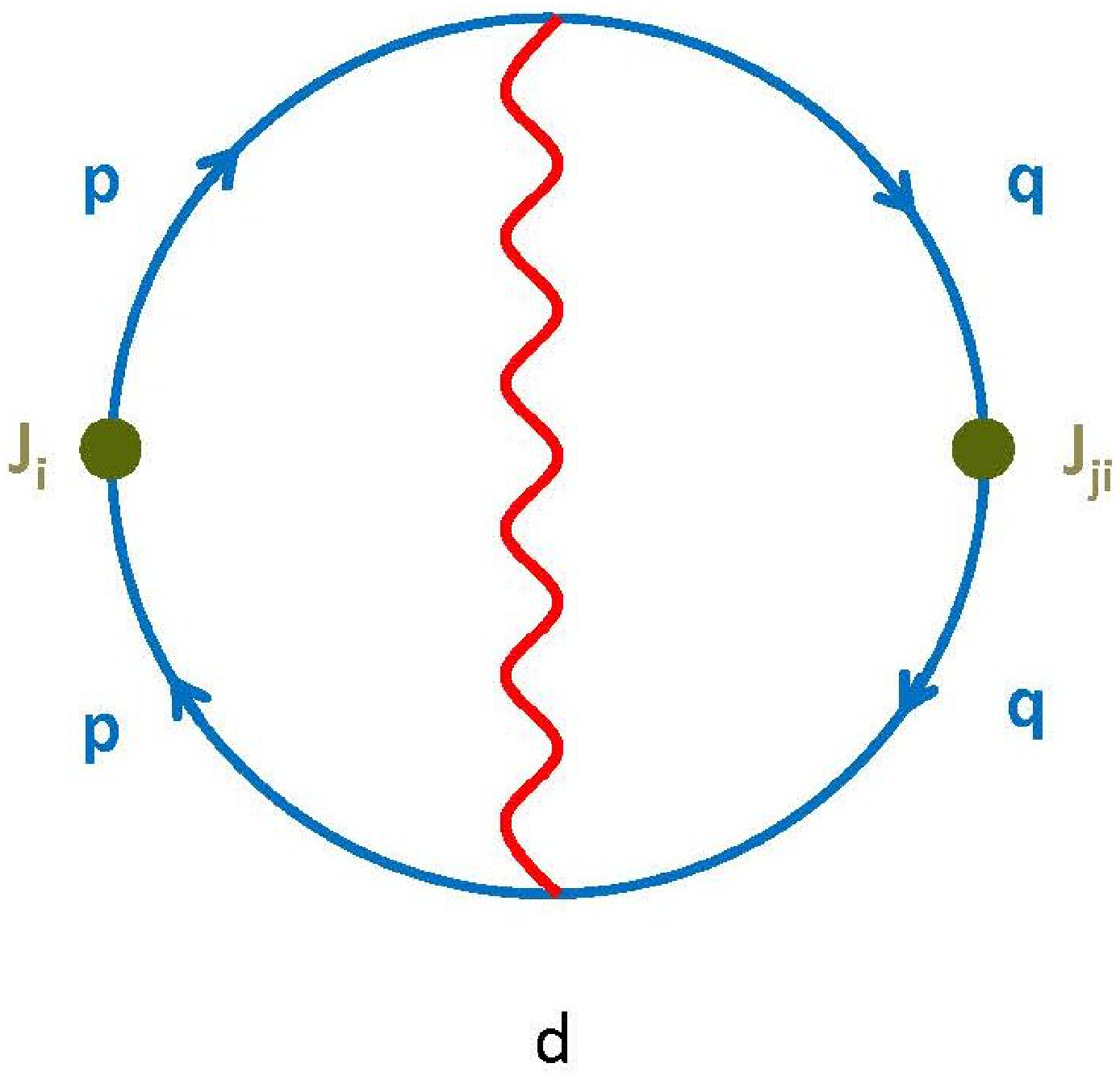}
\vskip0.6cm\hskip -12.0cm
\includegraphics[width=0.20\textwidth]{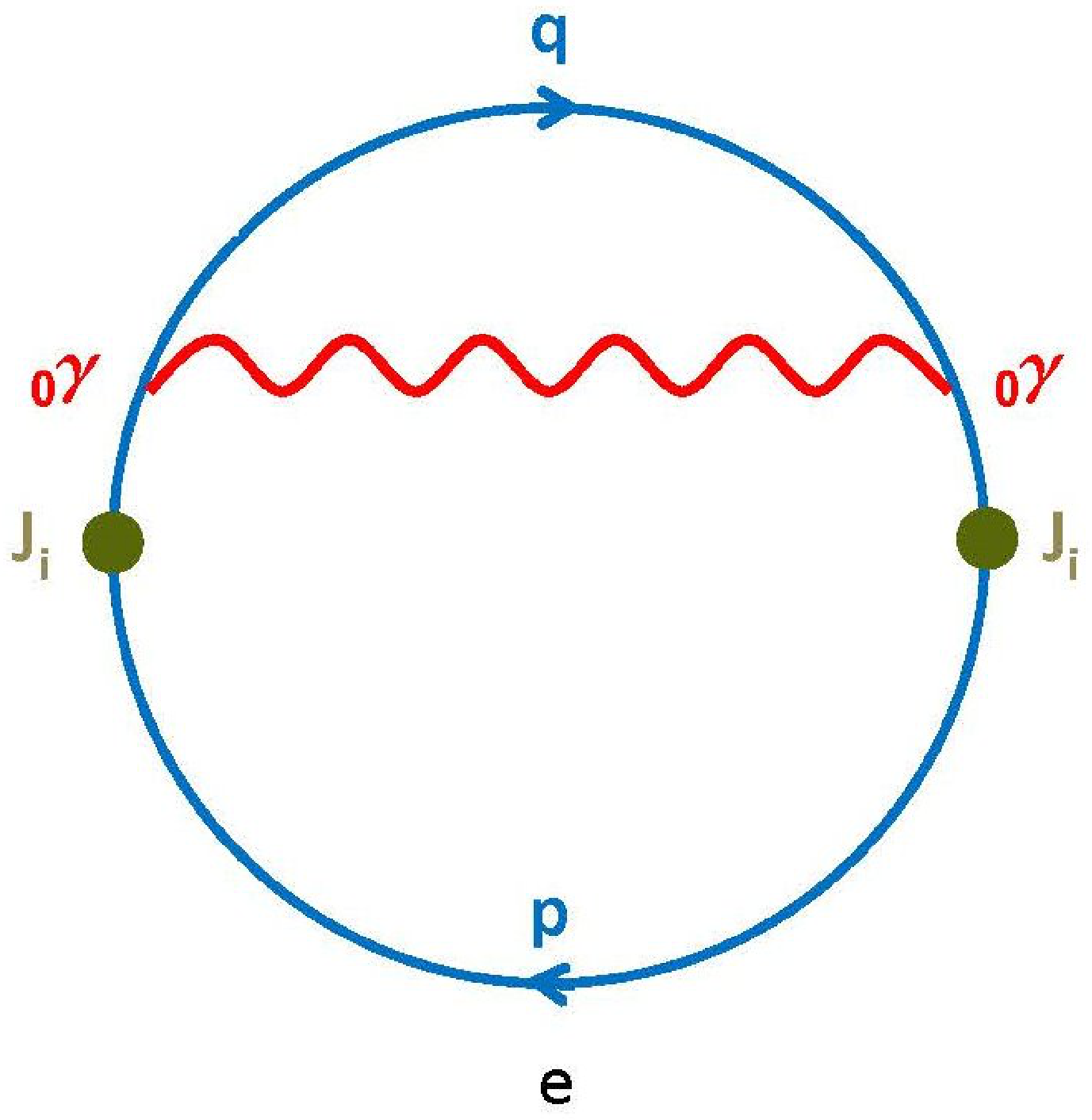}
\vskip-3.4cm\hskip 15.0cm
\includegraphics[width=0.40\textwidth]{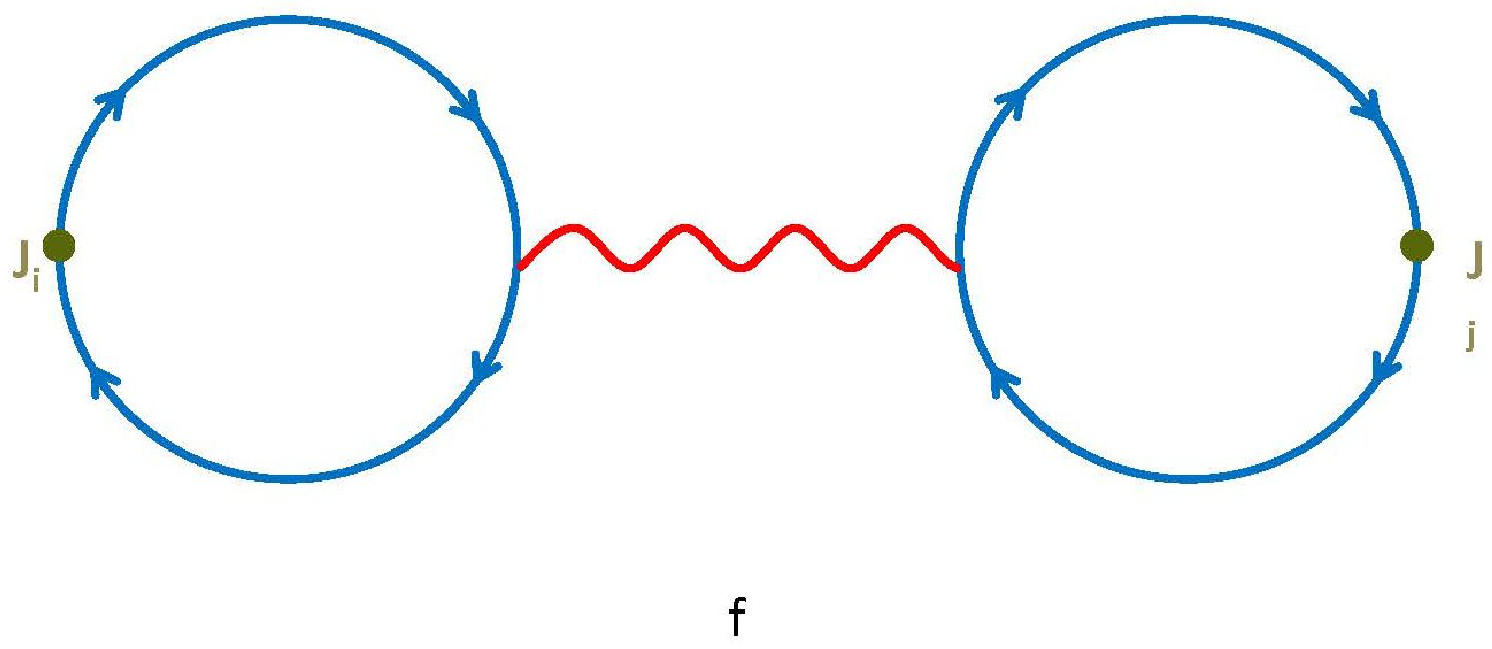}
\vskip -0.5cm
\caption{The tight binding model and Feynman diagrams for conductivity tensor.\hfil\break
a. The cubic lattice.\hfil\break
b. Propagators and Coulomb interaction.\hfil\break
c. The leading order conductivity.\hfil\break
d. The vertex correction.\hfil\break
e. The self energy correction.\hfil\break
f. The "glasses" diagram.}
\end{figure*}

\begin{figure*}
\hskip -12.0cm
\includegraphics[width=0.30\textwidth]{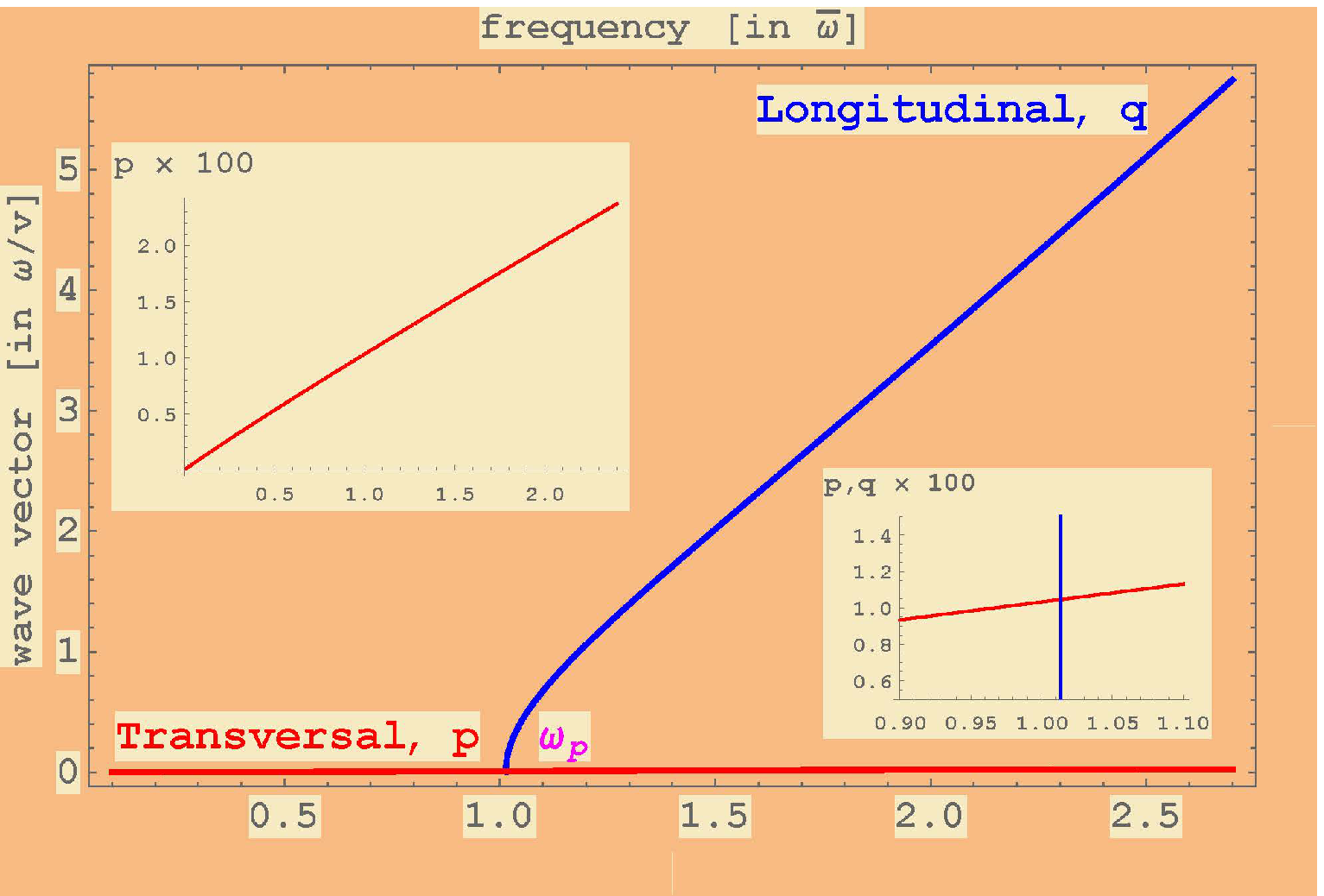}
\vskip -4.0cm\hskip 50.0cm
\includegraphics[width=0.31\textwidth]{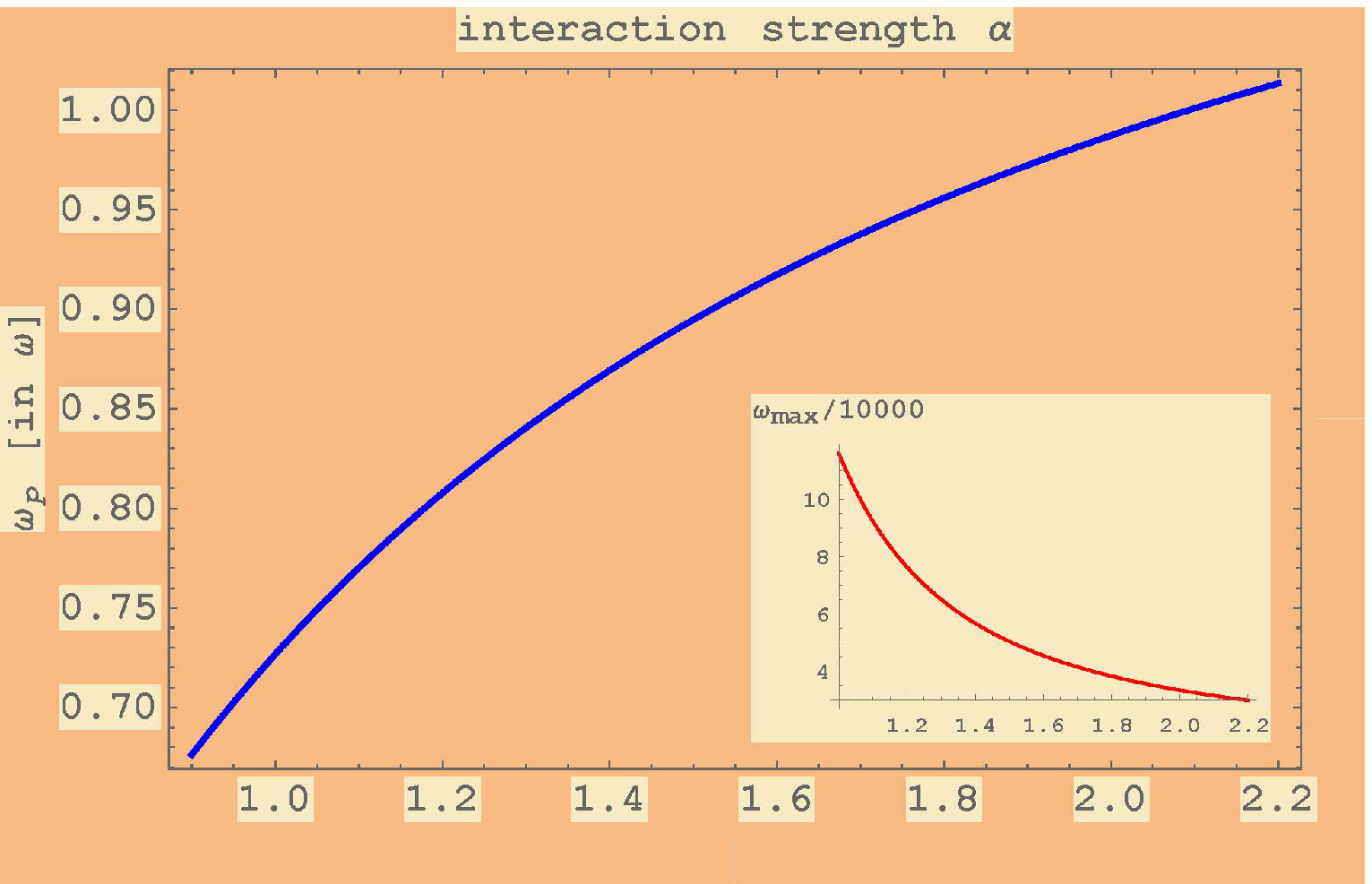}
\caption{Two branches of excitations in neutral plasma of the isotropic Weyl
semimetal.\hfil\break
a. The transverse (red) and longitudinal (blue) wave vectors are
monotonically increasing as function of frequency. At the intersection point
at $\left\{ q_{m},\omega _{m}\right\} $ the electromagnetic wave is
monochromatic (lower right insert).\hfil\break
b. The lower limiting (plasma) frequency at which the longitudinal wave
appears as a function of the interaction strength $\alpha $. The insert
shows the upper limiting frequency.}
\end{figure*}

\begin{figure*}
\vskip-0.5cm\hspace{-12.0cm} 
\includegraphics[width=0.35\textwidth]{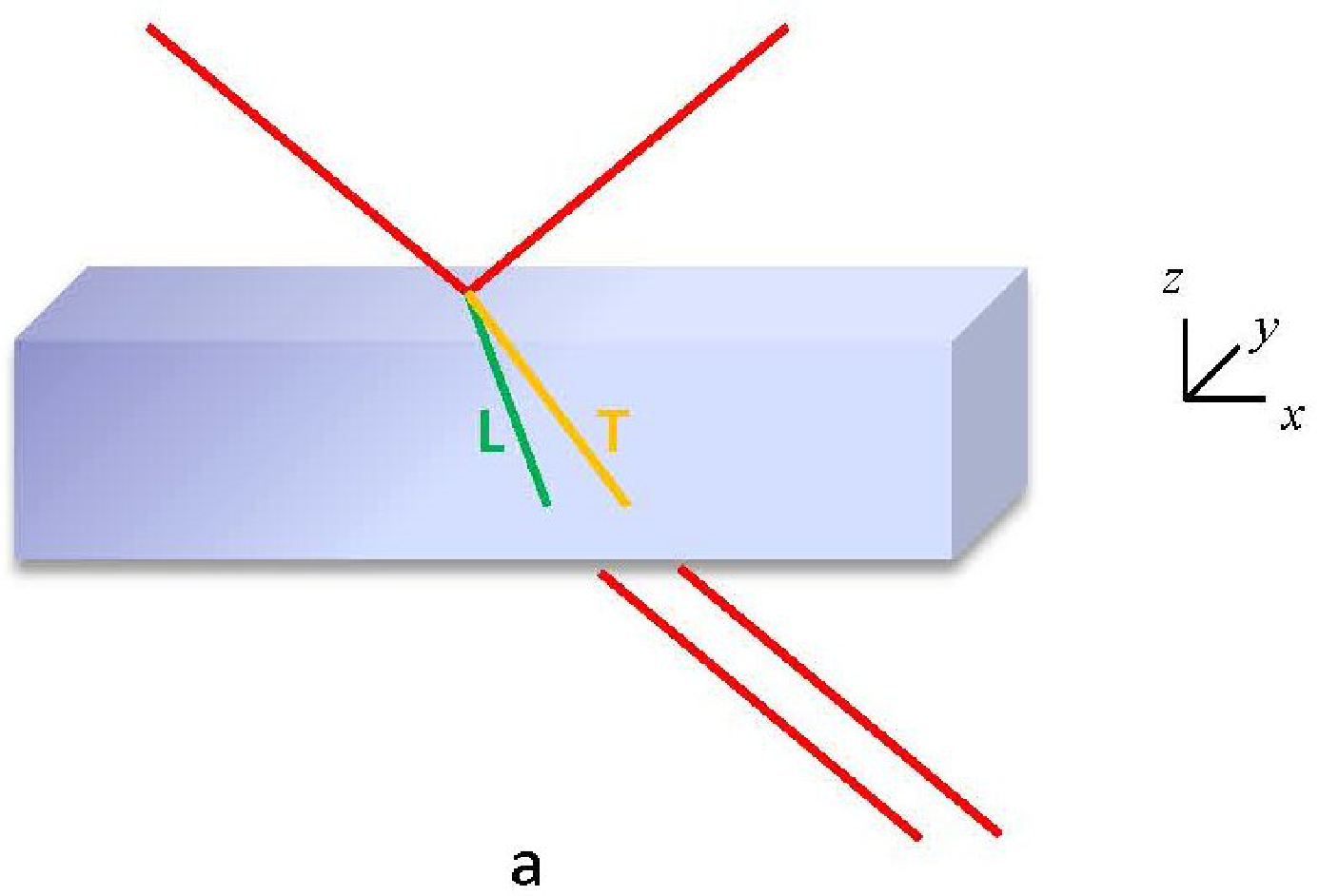}
\vskip-6.0cm\hspace{15.0cm}
\includegraphics[width=0.35\textwidth]{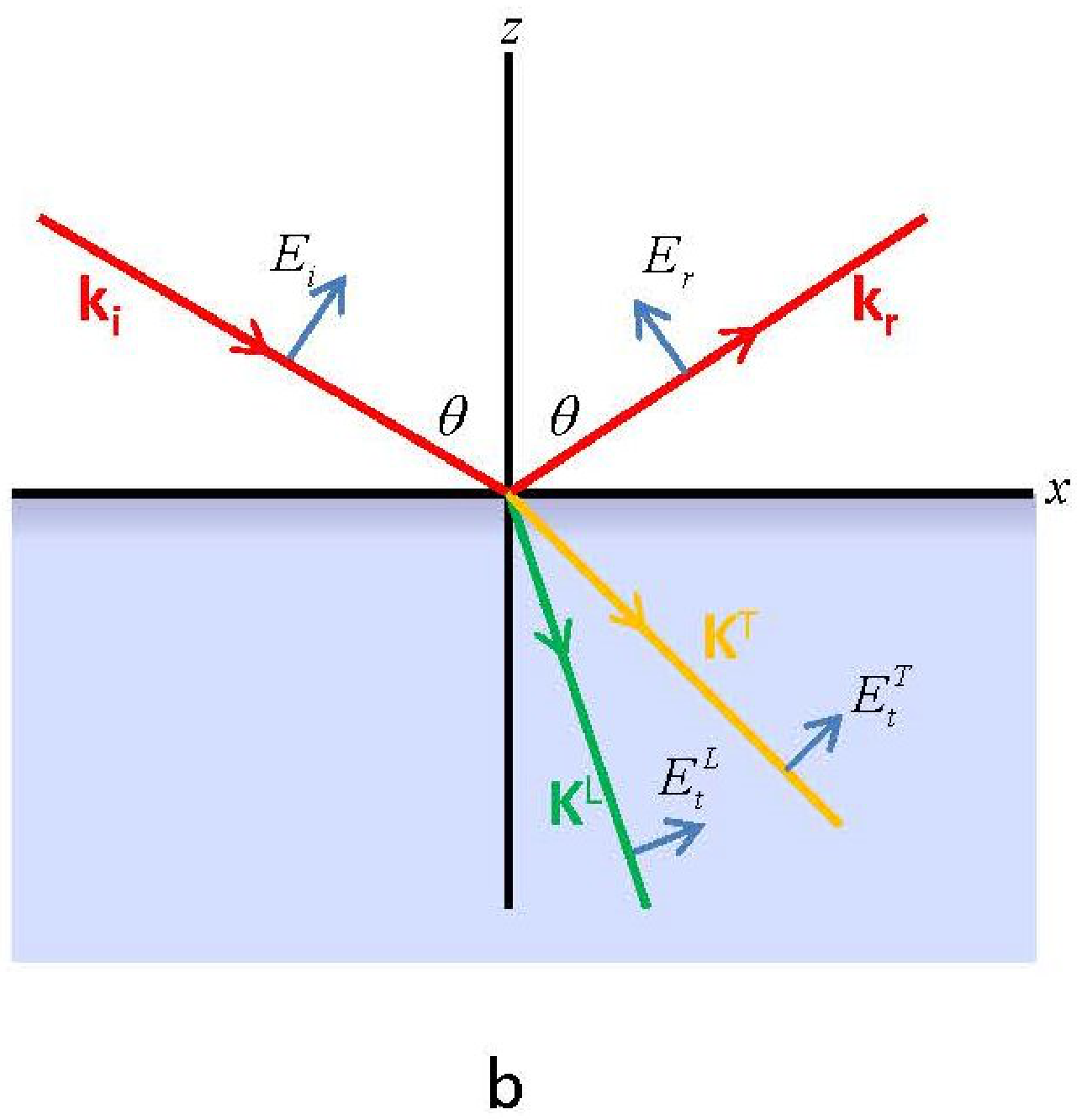}
\caption{Transmission and reflection of light in Weyl semimetal.\hfil\break
a. The light beam is either reflected or transmitted inside the WSM, where
it splits into a longitudinal (green) and a transverse (yellow) mode.
\hfil\break
b. The p-polarized light incident monochromatic beam with wave vector $%
\mathbf{k}$ incident at angle $\theta $ is split in the WSM into
longitudinal (green) and transverse (yellow) modes with wave vectors $%
\mathbf{q}$ and $\mathbf{p}$, respectively. Directions of the electric
fields are shown in blue.}
\end{figure*}

\begin{figure*}
\hskip -12.0cm
\includegraphics[width=0.30\textwidth]{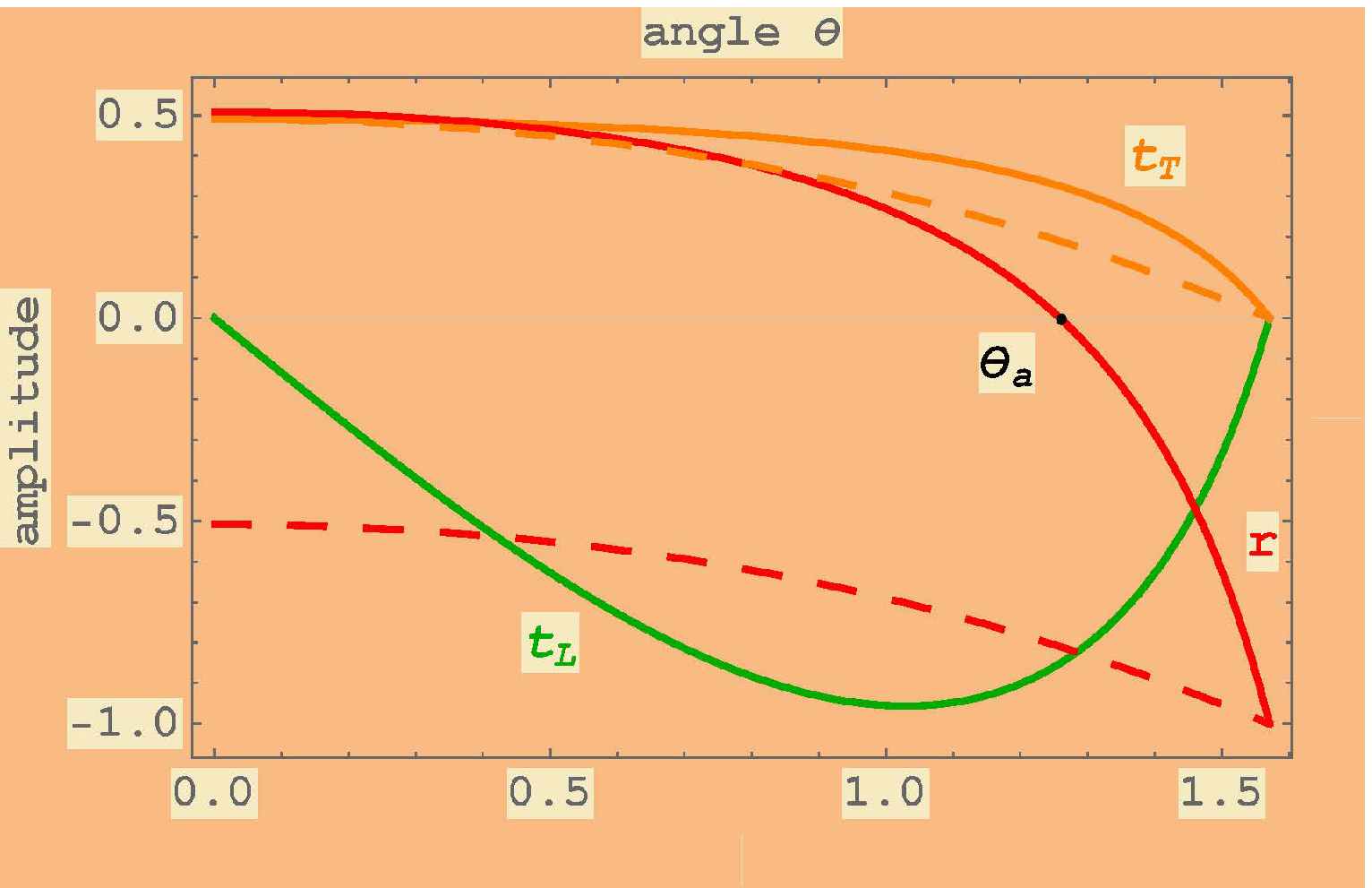}
\vskip -3.9cm\hskip 50.0cm
\includegraphics[width=0.31\textwidth]{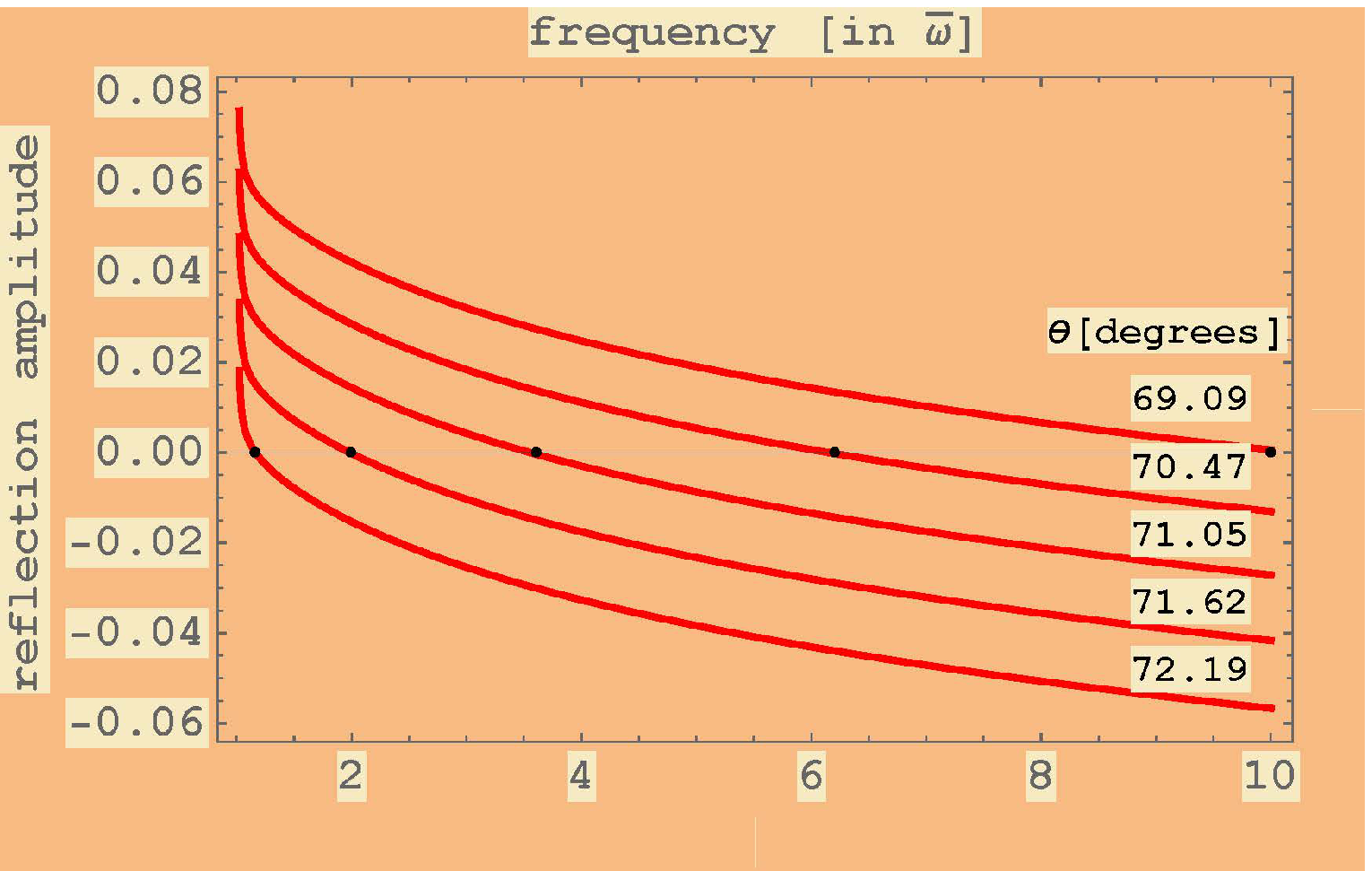}
\caption{Transmission and reflection of light in Weyl semimetal.\hfil\break
a. Reflection and transmission amplitudes as function of the incident angle 
$\theta $. The p-polarization amplitudes (solid): reflected (red),
transmitted transversal (yellow) and longtudinal (green); the incident
radiation is totally absorbed at the interface at angle $\theta _{a}$. The
s-polarization amplitudes are plotted dashed.
\hfil\break
b. The (p) reflection amplitude $r$ as function of frequency for various
incident angles displaying the total absorption.}
\end{figure*}

\begin{figure*}
\hskip -12.0cm
\includegraphics[width=0.35\textwidth]{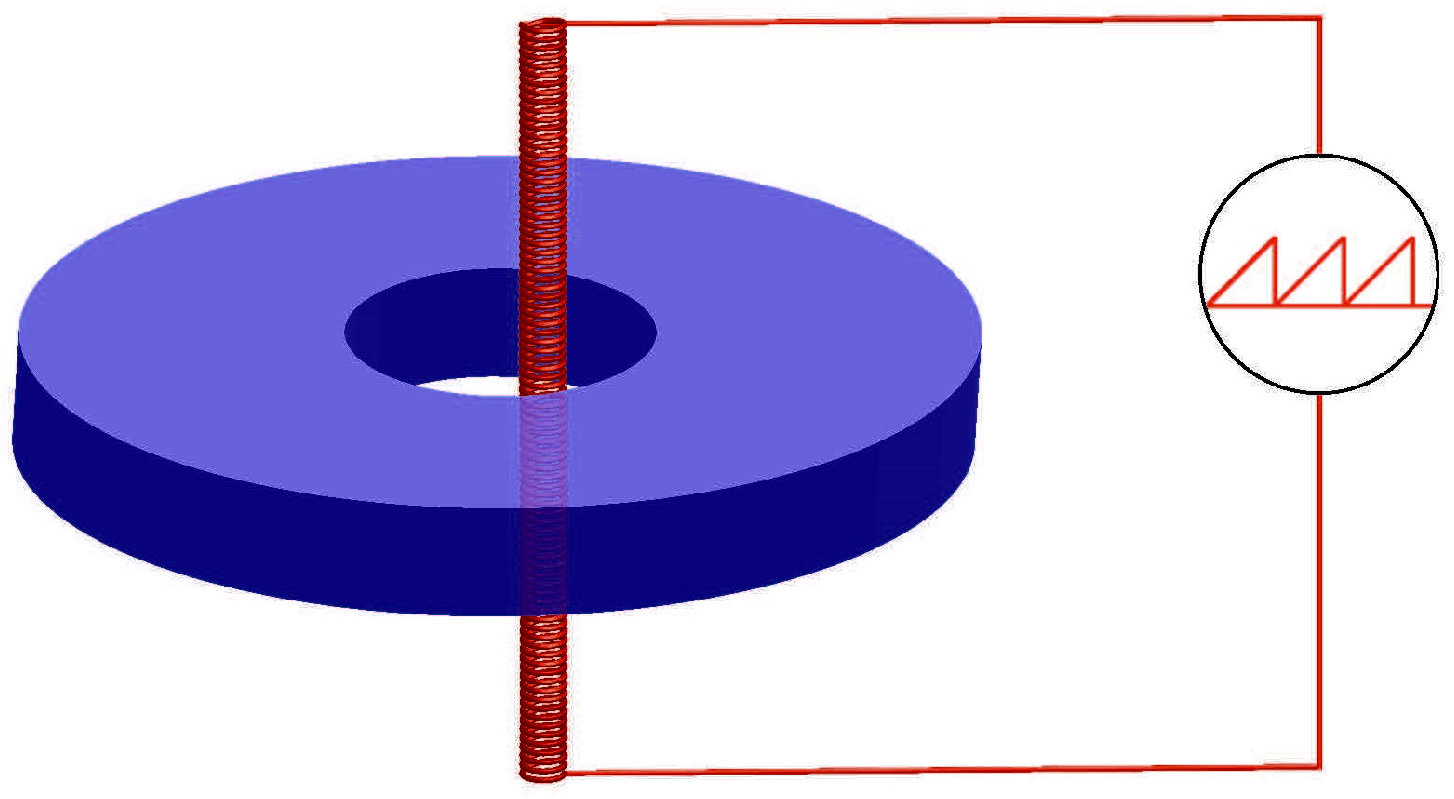}
\vskip -5.5cm\hskip 50.0cm
\includegraphics[width=0.30\textwidth]{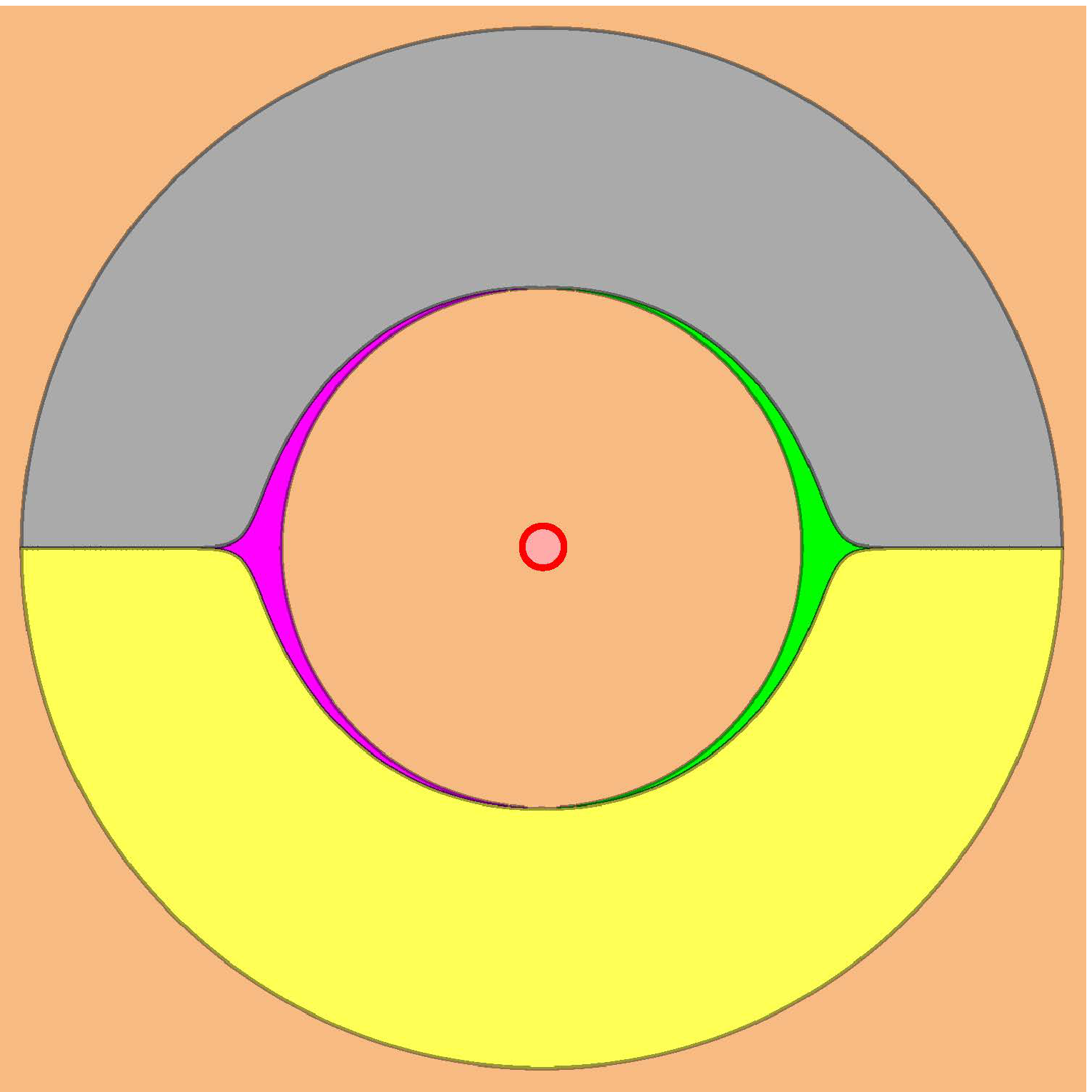}
\caption{Transmission and reflection of light in Weyl semimetal.\hfil\break
a. A thin solenoid inserted at distance $d$ from the center of a hollow Weyl
semimetal cylinder demonstrates the charging effect. The solenoid generates
pulses of magnetic flux. During the pulse the internal surface of the
aperture of radius $R$ gets charged.
\hfil\break
b. Charging effect in a system combining WSM and a conductor. The upper
segment is a WSM, while the lower is an ordinary 'local' conductor (a
semiconductor or a metal). The solenoid is placed in the center of the
aperture. Charges appear both on the internal surface and interfaces. The
surface charge density \ref{Q2} as function of polar angle with $d/R=1/2$.
}
\end{figure*}

\end{document}


\part{Supplemental Information 1}

\section{Calculation of the interaction corrections to conductivity of WSM}

\subsection{The current density in the tight binding model}

The minimal substitution determines the coupling of the external
electromagnetic field, described by vector potential $A_{i}$ with electrons
on the lattice:%
\begin{equation}
\widehat{K}_{mc}\left[ A\left( r,t\right) \right] =\frac{i}{2}%
\dsum\limits_{n,i}\Gamma _{\mathbf{n},i}c_{\mathbf{n}}^{\alpha \dagger
}\sigma _{i}^{\alpha \beta }c_{\mathbf{n}+\mathbf{a}_{i}}^{\beta }+hc, 
\tag{SI1(1)}
\end{equation}%
where the hopping integral $\Gamma _{\mathbf{n},i}$ becomes 
\begin{equation}
\Gamma _{\mathbf{n},i}=\gamma \exp \left\{ i\frac{ea}{c\hbar }%
\int_{s=0}^{1}A_{i}\left( \mathbf{n}+s\mathbf{a}_{i},t\right) \right\} \text{%
.}  \tag{SI1(2)}
\end{equation}%
Applying the charge symmetry transformation $c_{n}^{\alpha }\rightarrow
e^{i\chi _{n}\left( t\right) }c_{n}^{\alpha }$ the current density on the
link is 
\begin{equation}
J_{\mathbf{n}i}\left( t\right) \equiv -c\frac{\delta }{\delta A_{\mathbf{n}%
i}\left( t\right) }\widehat{K}_{mc}.  \tag{SI1(3)}
\end{equation}%
In linear response the current density operator is expanded up to the first
order in $A_{\mathbf{n}i}$ as $J_{\mathbf{n}i}\left( t\right) $ $=J_{\mathbf{%
n}i}^{p}\left( t\right) +J_{\mathbf{n}i}^{d}\left( t\right) $:

\begin{eqnarray}
J_{\mathbf{n}i}^{p}\left( t\right) &=&\frac{ev}{2\hbar }c_{\mathbf{n}%
}^{\alpha \dagger }\sigma _{i}^{\alpha \beta }c_{\mathbf{n}+\mathbf{a}%
_{i}}^{\beta }+hc  \nonumber \\
J_{\mathbf{n}i}^{d}\left( t\right) &=&i\frac{e^{2}va}{2c\hbar ^{2}}%
\int_{s=0}^{1}A_{i}\left( \mathbf{n}+s\mathbf{a}_{i},t\right) c_{\mathbf{n}%
}^{\alpha \dagger }\sigma _{i}^{\alpha \beta }c_{\mathbf{n}+\mathbf{a}%
_{i}}^{\beta }+hc,  \TCItag{SI1(4)}
\end{eqnarray}%
Normalized operators in momentum space are $c_{\mathbf{n}}^{\alpha }=%
\mathcal{N}^{-1/2}\dsum\limits_{\mathbf{k}}e^{-i\mathbf{k}\cdot \mathbf{n}%
}c_{\mathbf{k}}^{\alpha }$, where the number of unit cells is $\mathcal{N}=%
\mathcal{V}/a^{3}$ with $\mathcal{V}=a^{3}\dsum\nolimits_{\mathbf{n}}1$,
where $\widehat{k}_{i}=\sin \left( ak_{i}\right) $. There are eight Weyl
points at which $\varepsilon _{\mathbf{k}}=\left\vert \widehat{\mathbf{k}}%
\right\vert =0$ inside the Brillouin zone. The calculation therefore was
performed for $N=8$.

\subsection{Matsubara Action and Feynman rules}

It is useful to represent the electron gas via the Matsubara action $\left(
\tau =it\right) $ involving the $3+1$ Fermion field $\psi $ and the static
photon (auxiliary) field $\phi :$

\begin{equation}
A=\dsum\limits_{\mathbf{n}}\int_{\tau }\left\{ 
\begin{array}{c}
\psi _{\mathbf{n}}^{\dagger }\left( \tau \right) \left( \frac{d}{d\tau }%
-ie\phi _{\mathbf{n}}\right) \psi _{\mathbf{n}}\left( \tau \right) -\frac{1}{%
2}\dsum\limits_{i}\left( \psi _{\mathbf{n}}^{\dagger }\left( \tau \right)
\sigma _{i}\psi _{\mathbf{n}+\mathbf{a}_{i}}\left( \tau \right) +hc\right)
\\ 
+\frac{1}{2}\dsum\limits_{i}\left( 2\phi _{\mathbf{n}}\left( \tau \right)
-\phi _{\mathbf{n}+\mathbf{a}_{i}}\left( \tau \right) -\phi _{\mathbf{n}-%
\mathbf{a}_{i}}\left( \tau \right) \right) \phi _{\mathbf{n}}\left( \tau
\right)%
\end{array}%
\right\} .  \tag{SI1(5)}
\end{equation}%
We are using convenient units where $a=v=\hbar =1.$ The Feynman rules can be
read from the Fourier transform 
\begin{equation}
A=\dsum\limits_{\mathbf{k}\omega }\left\{ \psi _{\mathbf{k}\omega }^{\dagger
}\left( i\omega -\widehat{\mathbf{k}}\cdot \sigma \right) \psi _{\mathbf{k}%
\omega }-ie\dsum\limits_{\mathbf{p}\nu }\psi _{\mathbf{k}\omega }^{\dagger
}\psi _{\mathbf{k}-\mathbf{p},\omega -\nu }\phi _{\mathbf{p}\nu }\right\} +%
\frac{1}{2}\dsum\limits_{\mathbf{p}\nu i}\phi _{\mathbf{p}\nu }^{\ast
}\left( 1-\cos \left( p_{i}\right) \right) \phi _{\mathbf{p}\nu }. 
\tag{SI1(6)}
\end{equation}

The fermion propagator is,%
\begin{equation}
G_{\mathbf{p}\omega }=\left( i\omega -\widehat{\mathbf{p}}\cdot \sigma
\right) ^{-1}=\frac{-i\omega -\widehat{\mathbf{p}}\cdot \sigma }{\omega
^{2}+\varepsilon _{\mathbf{p}}^{2}},  \tag{SI1(7)}
\end{equation}%
while the photon propagator is frequency independent, $g_{\mathbf{p\omega }%
}=v_{\mathbf{p}}$. The vertex is: $ie\delta _{\mathbf{k}-\mathbf{k}^{\prime
}+\mathbf{p}}\delta _{\omega -\omega ^{\prime }}$. These Feynman rules are
used to calculate the linear response diagramatically.

\section{Direct calculation of the conductivity tensor}

\subsection{General relation between conductivity tensor and the dielectric
constant}

We calculate the transversal and longitudinal conductivities $\sigma _{T}$
and $\sigma _{L}$. From Eq.(3) we derive expression for two scalars formed
from the conductivity tensor $\sigma _{ij}:$

\begin{eqnarray}
2\sigma _{T}\left( \omega ,k\right) +\sigma _{L}\left( \omega ,k\right)
&=&\sigma _{ii}\left( \omega \right)  \TCItag{SI1(8)} \\
\sigma _{L}\left( \omega ,k\right) &=&\lim_{k\rightarrow 0}\frac{k_{i}k_{j}}{%
k^{2}}\sigma _{ij}\left( \omega \right) .  \nonumber
\end{eqnarray}%
Of course, due to charge conservation the transverse conductivity is
proportional to the dielectric constant (Matsubara)[1]

\begin{equation}
\sigma _{L}\left( \omega ,k\right) =\frac{\omega }{4\pi }\left(
1-\varepsilon \left( \omega ,k\right) \right) .  \tag{SI1(9)}
\end{equation}

\subsection{Non-interacting WSM}

The ac conductivity of the non-interacting WSM on the lattice is obtained
from the Kubo formula. The trace is (for positive $\omega $)

\begin{equation}
\sigma _{ii}\left( \omega \right) =\frac{1}{\omega }e^{2}\dsum\limits_{%
\mathbf{p},\nu }\text{Tr}\left\{ \sigma _{i}G_{\mathbf{p},\nu }\sigma
_{i}\left( G_{\mathbf{p},\nu +\omega }-G_{\mathbf{p},\nu }\right) \right\}
=2e^{2}\dsum\limits_{\mathbf{p}}\frac{\omega }{\varepsilon _{\mathbf{p}%
}\left( \omega ^{2}+4\varepsilon _{\mathbf{p}}^{2}\right) },  \tag{SI1(10)}
\end{equation}%
while the longitudinal conductivity is

\begin{eqnarray}
\sigma _{L}\left( \omega \right)  &=&\lim_{k\rightarrow 0}\frac{e^{2}}{%
\omega k^{2}}\dsum\limits_{\mathbf{p},\nu }\text{Tr}\left\{ \mathbf{k}\cdot
\sigma G_{\mathbf{p},\nu }\mathbf{k}\cdot \sigma \left( G_{\mathbf{p},\nu
+\omega }-G_{\mathbf{p},\nu }\right) \right\}   \TCItag{SI1(11)} \\
&=&\frac{2}{3}e^{2}\dsum\limits_{\mathbf{p}}\frac{\omega }{\varepsilon _{%
\mathbf{p}}\left( \omega ^{2}+4\varepsilon _{\mathbf{p}}^{2}\right) }=\frac{1%
}{3}\sigma _{ii}\text{.}  \nonumber
\end{eqnarray}%
From here one deduces that $\sigma _{T}=\sigma _{L}$ and thus no nonlocal
conductivity arises in the non-interacting relativistic system, see general
arguments in [27].

On the lattice one obtains%
\begin{equation}
\sigma _{T}^{\left( 0\right) }=\sigma _{L}^{\left( 0\right) }=\sigma _{0}=%
\frac{e^{2}}{v\hbar }\omega \left( -\frac{7}{32\pi }+\frac{2}{3\pi ^{2}}\log
\left( \omega \right) \right) .  \tag{SI1(12)}
\end{equation}%
This result is consistent with [20]. The dielectric constant in leading
order is 
\begin{equation}
\varepsilon ^{\left( 0\right) }=e^{2}\left( \frac{7}{4}-\frac{8}{3\pi }\log
\left( \omega \right) \right) .  \tag{SI1(13)}
\end{equation}

\subsection{The interaction corrections:}

\subsubsection{The trace in Eq.(SI1(8))}

The three interaction corrections to the trace of the conductivity tensor
are the self energy, the vertex and the 'glasses' corrections, see Figs.2d-f,

\begin{eqnarray}
\sigma _{ii}^{ver} &=&\frac{e^{4}}{\omega }\tsum\limits_{\mathbf{pq}\nu \rho
}\text{Tr}\left[ v_{\mathbf{q}-\mathbf{p}}\text{ }\widetilde{\mathbf{p}}%
\cdot \widetilde{\mathbf{q}}\text{ }G_{\mathbf{p},\nu }G_{\mathbf{q},\rho
}\sigma _{i}G_{\mathbf{q},\rho +\omega }G_{\mathbf{p},\omega +\nu }\sigma
_{i}\right] ;  \nonumber \\
\sigma _{ii}^{se} &=&-\frac{2e^{4}}{\omega }\tsum\limits_{\mathbf{pq}\nu
\rho }\text{Tr}\left[ v_{\mathbf{q}-\mathbf{p}}\widetilde{\mathbf{p}}\cdot 
\widetilde{\mathbf{p}}\text{ }G_{\mathbf{p},\nu }G_{\mathbf{q},\rho }G_{%
\mathbf{p},\nu }\sigma _{i}G_{\mathbf{p},\omega +\nu }\sigma _{i}\right] ; 
\TCItag{SI1(14)} \\
\sigma _{ii}^{gl} &=&\frac{e^{4}}{\omega }v_{\mathbf{k}}\left( \tsum\limits_{%
\mathbf{p},\nu }\text{Tr}G_{\mathbf{p},\nu }G_{\mathbf{k}+\mathbf{p},\omega
+\nu }\right) ^{2}\text{,}  \nonumber
\end{eqnarray}%
where the first term is the vertex, the second term the self energy and the
third term the 'glasses' diagram. The following shorthand is useful:%
\begin{eqnarray*}
\widehat{k_{i}} &\equiv &\sin k_{i};\text{ }\widetilde{k_{i}}\equiv \cos
\left( k_{i}\right) ;\text{ }\widehat{2k_{i}}\equiv \sin \left(
2k_{i}\right) ;\varepsilon _{k}=\sqrt{\widehat{\mathbf{k}}^{2}}; \\
\widehat{\mathbf{k}}^{n} &=&\left\{ \widehat{k_{1}}^{n},\widehat{k_{2}}^{n},%
\widehat{k_{3}}^{n}\right\} \text{, \ }\widetilde{\mathbf{k}}^{n}=\left\{ 
\widetilde{k_{1}}^{n},\widetilde{k_{2}}^{n},\widetilde{k_{3}}^{n}\right\} 
\text{.}
\end{eqnarray*}

Calculating the trace and integrating over the internal frequencies $\nu $
and $\rho $ results in the following expressions for the three diagrams
(symmetrized in $\mathbf{p}$ and $\mathbf{q}$). The vertex

\begin{equation}
\sigma _{ii}^{ver}=\frac{e^{4}}{\omega }\frac{N}{288}\tsum\limits_{\mathbf{qp%
}}\frac{v_{\mathbf{q-p}}\left( \Xi _{1}+\Xi _{2}\right) }{\varepsilon _{%
\mathbf{p}}\varepsilon _{\mathbf{q}}\left( \omega ^{2}+4\varepsilon _{%
\mathbf{p}}^{2}\right) ^{2}\left( \omega ^{2}+4\varepsilon _{\mathbf{q}%
}^{2}\right) ^{2}}\text{,}  \tag{SI1(15)}
\end{equation}%
with%
\begin{eqnarray*}
\Xi _{1} &=&-\omega ^{2}\left\{ 4\text{ }\left( \widehat{\mathbf{p}}\cdot 
\widehat{\mathbf{q}}\right) \text{ }\left( \widetilde{\mathbf{p}}\cdot 
\widetilde{\mathbf{q}}\right) -\widehat{\mathbf{2p}}\cdot \widehat{\mathbf{2q%
}}\right\} ; \\
\Xi _{2} &=&2\left\{ 
\begin{array}{c}
\left( \widehat{\mathbf{p}}\cdot \widehat{\mathbf{q}}\right) \text{ }\left( 
\widehat{\mathbf{2p}}\cdot \widehat{\mathbf{2q}}\right) +12\widetilde{%
\mathbf{p}}\cdot \widetilde{\mathbf{q}}-16\varepsilon _{\mathbf{p}}^{2}\text{
}\widetilde{\mathbf{p}}\cdot \widetilde{\mathbf{q}}+24\widetilde{\mathbf{q}}%
^{3}\cdot \widetilde{\mathbf{p}} \\ 
+4\left( \widetilde{\mathbf{p}}\cdot \widetilde{\mathbf{p}}\right) \text{ }%
\left( \widetilde{\mathbf{q}}\cdot \widetilde{\mathbf{q}}\right) \left( 
\text{ }\widetilde{\mathbf{p}}\cdot \widetilde{\mathbf{q}}\right) -8\left( 
\widetilde{\mathbf{p}}\cdot \widetilde{\mathbf{p}}\right) \text{ }\left( 
\widetilde{\mathbf{q}}^{3}\cdot \widetilde{\mathbf{p}}\right) +\mathbf{p}%
\longleftrightarrow \mathbf{q}%
\end{array}%
\right\} \text{,}
\end{eqnarray*}%
the self energy,

\begin{equation}
\sigma _{ii}^{se}=-\frac{e^{4}}{\omega }\frac{N}{24}\tsum\limits_{\mathbf{qp}%
}\frac{v_{\mathbf{q-p}}}{\varepsilon _{p}\varepsilon _{q}}\left( \frac{\Xi
_{3}+\Xi _{4}}{\varepsilon _{p}^{2}\left( \omega ^{2}+4\varepsilon
_{p}^{2}\right) ^{2}}+\mathbf{p}\longleftrightarrow \mathbf{q}\right) \text{,%
}  \tag{SI1(16)}
\end{equation}%
\begin{eqnarray*}
\Xi _{3} &=&\omega ^{2}\left\{ -\varepsilon _{p}^{4}\left( \widehat{\mathbf{p%
}}\cdot \widehat{\mathbf{q}}\right) +2\varepsilon _{p}^{2}\left( \widehat{%
\mathbf{p}}^{3}\cdot \widehat{\mathbf{q}}\right) +2\varepsilon
_{p}^{2}\left( \widehat{\mathbf{p}}\cdot \widehat{\mathbf{q}}\right) -\left( 
\widehat{\mathbf{p}}^{2}\cdot \widehat{\mathbf{p}}^{2}\right) \left( 
\widehat{\mathbf{p}}\cdot \widehat{\mathbf{q}}\right) \right\} \text{;} \\
\Xi _{4} &=&4\varepsilon _{p}^{2}\left\{ \varepsilon _{p}^{4}\left( \widehat{%
\mathbf{p}}\cdot \widehat{\mathbf{q}}\right) +2\varepsilon _{p}^{2}\left( 
\widehat{\mathbf{p}}^{3}\cdot \widehat{\mathbf{q}}\right) -2\varepsilon
_{p}^{2}\left( \widehat{\mathbf{p}}\cdot \widehat{\mathbf{q}}\right)
-3\left( \widehat{\mathbf{p}}^{2}\cdot \widehat{\mathbf{p}}^{2}\right)
\left( \widehat{\mathbf{p}}\cdot \widehat{\mathbf{q}}\right) \right\} \text{,%
}
\end{eqnarray*}%
and the "glasses" diagram that proportional to the square of the number of
the Dirac points points,

\begin{equation}
\sigma _{ii}^{gl}=\frac{e^{4}}{\omega }\frac{4\pi N}{3}\left(
\tsum\limits_{p}\frac{\widehat{p}_{y}^{2}\widetilde{p}_{x}}{\varepsilon _{%
\mathbf{p}}^{3}\left( \omega ^{2}+4\varepsilon _{\mathbf{p}}^{2}\right) }%
\right) ^{2}\text{.}  \tag{SI1(17)}
\end{equation}

The integration over the momenta is performed with the sum of the vertex and
the self energy diagrams, which ensures convergence. The integral for the
glasses diagram vanishes.

\begin{equation}
\sigma _{ii}^{\left( 1\right) }=-e^{4}\omega \left( \frac{1}{8}+\frac{1.308}{%
\pi }\log \left( \omega \right) +\frac{1}{3\pi ^{2}}\log ^{2}\left( \omega
\right) \right)  \tag{SI1(18)}
\end{equation}%
which expressed in physical units and physical frequency is

\[
\sigma _{ii}^{\left( 1\right) }=-\frac{e^{2}\alpha N}{\hbar v8}\omega \left[
-\frac{1.308}{2}-\frac{1}{3\pi }\log \left( \frac{a\omega }{v}\right)
+i\left( \frac{1}{24}+\frac{1.308}{\pi }\log \left( \frac{a\omega }{v}%
\right) +\frac{1}{3\pi ^{2}}\log ^{2}\left( \frac{a\omega }{v}\right)
\right) \right] 
\]

\subsubsection{The longitudinal conductivity via the dielectric constant,
Eq.(SI1(9)):}

The general expressions for the dielectric function via $\rho -\rho $
correlator can be written as $\varepsilon =1+e^{2}\varepsilon
^{1}+e^{4}\varepsilon ^{2}$ with

\begin{equation}
\varepsilon ^{1}=-e^{2}v_{\mathbf{k}}Tr\tsum\limits_{\mathbf{p}\nu }G_{%
\mathbf{p},\nu }G_{\mathbf{p}+\mathbf{k},\nu +\omega }  \tag{SI1(18)}
\end{equation}

and $\varepsilon ^{2}=\varepsilon ^{ver}+\varepsilon ^{se}+\varepsilon ^{gl}$
where

\begin{eqnarray}
\varepsilon ^{ver} &=&e^{4}v_{\mathbf{k}}\tsum\limits_{\mathbf{pq}\nu \rho }%
\text{Tr}G_{\mathbf{p}+\mathbf{k},\nu +\omega }G_{\mathbf{p},\nu }v_{\mathbf{%
q}}G_{\mathbf{p}+\mathbf{q},\rho }G_{\mathbf{p}+\mathbf{q}+\mathbf{k},\rho
+\omega }  \nonumber \\
\varepsilon ^{se} &=&-2e^{4}v_{\mathbf{k}}\tsum\limits_{\mathbf{pq}\nu \rho
}v_{\mathbf{q-p}}\text{Tr}G_{\mathbf{p},\nu }G_{\mathbf{q},\rho }G_{\mathbf{p%
},\nu }G_{\mathbf{p}+\mathbf{k},\omega +\nu }  \TCItag{SI1(19)} \\
\varepsilon ^{gl} &=&\left( e^{2}v_{\mathbf{k}}\text{Tr}\tsum\limits_{%
\mathbf{p}\nu }G_{\mathbf{p},\nu }G_{\mathbf{p}+\mathbf{k},\omega +\nu
}\right) ^{2}  \nonumber
\end{eqnarray}

The terms are expanded in powers of $k$, the trace is calculated and
integrated over the internal frequencies. All the $k^{0}$ and $k^{1}$ terms
vanish after integration over the internal frequencies.

\bigskip\ The relevant terms (symmetrized in $\mathbf{p}$ and $\mathbf{q}$):

\begin{equation}
\ \varepsilon ^{se}=-\frac{e^{4}\pi N}{6}\tsum\limits_{\mathbf{qp}}\frac{v_{%
\mathbf{q-p}}}{\varepsilon _{\mathbf{p}}\varepsilon _{\mathbf{q}}}\left\{ 
\begin{array}{c}
\left( \frac{\left( 3-\varepsilon _{\mathbf{p}}^{2}\right) \left( \omega
^{2}-4\varepsilon _{\mathbf{p}}^{2}\right) +2\left( \omega
^{2}+12\varepsilon _{\mathbf{p}}^{2}\right) }{2\left( \omega
^{2}+4\varepsilon _{\mathbf{p}}^{2}\right) ^{2}\varepsilon _{\mathbf{p}}^{2}}%
-\frac{\left[ 3\omega ^{2}+20\varepsilon _{\mathbf{p}}^{2}\right] \widehat{%
\left( \mathbf{2p}\right) }^{2}}{8\varepsilon _{\mathbf{p}}^{4}\left( \omega
^{2}+4\varepsilon _{\mathbf{p}}^{2}\right) ^{2}}\right) \widehat{\mathbf{p}}%
\cdot \widehat{\mathbf{q}} \\ 
-\frac{\left( \omega ^{2}+12\varepsilon _{\mathbf{p}}^{2}\right) \widehat{%
\mathbf{p}}^{3}\cdot \widehat{\mathbf{q}}}{\left( \omega ^{2}+4\varepsilon _{%
\mathbf{p}}^{2}\right) ^{2}\varepsilon _{\mathbf{p}}^{2}}+\mathbf{p}%
\longleftrightarrow \mathbf{q}%
\end{array}%
\right\}  \tag{SI1(20)}
\end{equation}

\begin{eqnarray}
\varepsilon ^{ver} &=&\frac{e^{4}\pi N}{48}\tsum\limits_{\mathbf{qp}}\frac{%
v_{\mathbf{q-p}}}{\varepsilon _{\mathbf{p}}^{3}\varepsilon _{\mathbf{q}%
}^{3}\left( \omega ^{2}+4\varepsilon _{\mathbf{p}}^{2}\right) \left( \omega
^{2}+4\varepsilon _{\mathbf{q}}^{2}\right) }  \TCItag{SI1(21)} \\
&&\left\{ 
\begin{array}{c}
\left[ 4\varepsilon _{\mathbf{p}}^{2}\varepsilon _{\mathbf{q}}^{2}-\omega
^{2}\left( \widehat{\mathbf{p}}\cdot \widehat{\mathbf{q}}\right) \right]
\left( \widehat{\mathbf{2p}}\cdot \widehat{\mathbf{2q}}\right) -2\varepsilon
_{\mathbf{p}}^{2}\left\{ 2\varepsilon _{\mathbf{q}}^{2}\left( \widehat{%
\mathbf{2p}}\cdot \widehat{\mathbf{2q}}\right) -\omega ^{2}\widehat{2\mathbf{%
q}}\cdot \left( \widehat{\mathbf{q}}\ast \widetilde{\mathbf{p}}\right)
\right\} \\ 
+4\varepsilon _{\mathbf{p}}^{2}\varepsilon _{\mathbf{q}}^{2}\left[ 4\left( 
\widehat{\mathbf{p}}\cdot \widehat{\mathbf{q}}\right) -\omega ^{2}\right]
\left( \widetilde{\mathbf{p}}\cdot \widetilde{\mathbf{q}}\right) +\mathbf{p}%
\longleftrightarrow \mathbf{q}%
\end{array}%
\right\}  \nonumber
\end{eqnarray}

\begin{equation}
\varepsilon ^{gl}=\frac{e^{4}\pi ^{2}N^{2}}{4}\left\{ \tsum\limits_{\mathbf{p%
}}\frac{-\varepsilon _{\mathbf{p}}^{2}+\frac{1}{3}\left( \widehat{\mathbf{p}}%
^{2}\cdot \widehat{\mathbf{p}}^{2}\right) +\frac{1}{12}\widehat{\mathbf{2p}}%
^{2}}{\varepsilon _{\mathbf{p}}^{3}\left( \omega ^{2}+4\varepsilon _{\mathbf{%
p}}^{2}\right) }\right\} ^{2}  \tag{SI1(22)}
\end{equation}

\bigskip

The numerical integration over the internal momenta results in (expressed in
physical units, Matsubara frequency)

\begin{equation}
\varepsilon =\frac{e^{2}}{\hbar v}\alpha \left\{ 
\begin{array}{c}
N\left[ -0.055+0.023\log \left( \frac{a\omega }{v}\right) -\frac{1}{18\pi }%
\log ^{2}\left( \frac{a\omega }{v}\right) \right] \\ 
+\left( \frac{N}{24}\right) ^{2}\left[ \frac{21}{32}-\frac{1}{\pi }\log
\left( \frac{a\omega }{v}\right) \right] ^{2}%
\end{array}%
\right\} \text{.}  \tag{SI1(23)}
\end{equation}

and in physical frequency:

\begin{equation}
\varepsilon =\frac{e^{2}}{\hbar v}\alpha \left\{ 
\begin{array}{c}
N\left[ -0.055+0.022\log \left( \frac{a\omega }{v}\right) -\frac{1}{18\pi }%
\log ^{2}\left( \frac{a\omega }{v}\right) +\frac{1}{18\pi }\frac{\pi ^{2}}{4}%
+i\left( 0.036-\frac{1}{18}\log \left( \frac{a\omega }{v}\right) \right) %
\right] \\ 
+\left( \frac{N}{24}\right) ^{2}\left[ \frac{21}{32}-\frac{1}{\pi }\log
\left( \frac{a\omega }{v}\right) -i\frac{1}{2}\right] ^{2}%
\end{array}%
\right\}  \tag{SI1(24)}
\end{equation}

\section{Supplemental Information 2. Surface charging}

\subsection{Corbino geometry with a fluxon out of the center}

Assume that the disk is thick enough so that the system is $z$ translational
invariant. A thin fluxon placed at distance $d$ on the $\mathbf{\hat{x}}$
axis. The flux therefore can be represented by $\Phi \left( \mathbf{r}%
\right) =\alpha t\delta \left( \mathbf{b}\right) ;\mathbf{b}=\mathbf{r}-d%
\mathbf{\hat{x}}$; see Fig.6. This produces the rotational electric field

\begin{equation}
E_{i}^{sol}\left( \mathbf{r}\right) =\frac{\alpha }{2\pi b^{2}}\varepsilon
_{ij}b_{j}.  \tag{SI2(1)}
\end{equation}

Surface charges are induced at the aperture with density $Q\left( \phi
\right) .$ The density is obtained from the condition that the normal
component of the current on the aperture surface on the WSM side vanishes, $%
\mathbf{n\cdot J=0}$. The current is caused by the solenoidal field with
conductivity $\sigma _{T}$ and the irrotational field (induced by the
surface charges) with conductivity $\sigma _{L}$. The irrational field
(denoting the distance from a point at the surface to the observation point $%
\mathbf{r}$ by $\mathbf{s}=\mathbf{r-}R\mathbf{n}^{\prime }$) is:

\begin{equation}
E_{i}^{irr}\left( \mathbf{r}\right) =R\int_{\phi =0}^{2\pi }\int_{z}\frac{%
s_{i}\text{ }Q\left( \phi \right) }{\left( z^{2}+s^{2}\right) ^{3/2}}%
=2R\int_{\phi =0}^{2\pi }\frac{\left( r_{i}\mathbf{-}Rn_{i}\right) Q\left(
\phi \right) }{s^{2}}.  \tag{SI2(2)}
\end{equation}

Inside the semi-metal near the surface one has for its normal component (see
Appendix)

\begin{equation}
\mathbf{n\cdot E}^{irr}\mathbf{=}2\pi Q\left( \phi \right) .  \tag{SI2(3)}
\end{equation}%
The above condition is then

\begin{equation}
\sigma _{T}\frac{\alpha n_{i}}{2\pi \left\vert R\mathbf{n}-d\mathbf{\hat{x}}%
\right\vert ^{2}}\varepsilon _{ij}\left( R\mathbf{n}-d\mathbf{\hat{x}}%
\right) _{j}+\sigma _{L}2\pi Q\left( \phi \right) =0,  \tag{SI2(4)}
\end{equation}%
so that one obtains Eq.(22)

\subsection{ The WSM - (Ohmic) conductor circuit}

The fluxon (in the $+z$ direction) is located in the center of a composite
cylinder of WSM with conductivities $\sigma _{T},\sigma _{L}$ in the 'upper' 
$\left( u\right) $ half $\left( 0<\phi <\pi \right) $ and a local conductor
with Ohmic conductivity $\sigma $ in the lower $\left( d\right) $ segment $%
\pi <\phi <2\pi .$ Surface charges are induced at the aperture surface and
at in the interface between the two materials. Their densities are
determined by the condition that the normal component of the current on the
aperture surface vanishes. Only the irrotational field is involved which has
contributions from the two interfaces and the surface charges. The
contribution from the circular surface is dominated by the local charge (see
Appendix). $\Sigma \left( x\right) $ denotes the interface charge density
and\ $Q\left( \phi \right) $ the aperture surface charge density.

In the upper segment, the WSM, the condition at the boundary is then (in the
lower segment the condition is trivially fulfilled)

\begin{equation}
J_{r}=\sigma _{L}E_{r}^{irr}\left( R,\phi \right) \mathbf{=}\sigma _{L}\left[
2\int_{x=R}^{\infty }\frac{\mathbf{r}}{R}\mathbf{\cdot }\frac{\left( \mathbf{%
r}-\left\{ x,0\right\} \right) \Sigma \left( x\right) }{\left( \mathbf{r}%
-\left\{ x,0\right\} \right) ^{2}}+2\int_{x=-\infty }^{-R}\frac{\mathbf{r}}{R%
}\mathbf{\cdot }\frac{\left( \mathbf{r}-\left\{ x,0\right\} \right) \Sigma
\left( x\right) }{\left( \mathbf{r}-\left\{ x,0\right\} \right) ^{2}}-2\pi
Q\left( \phi \right) \right] |_{r=R}=0,  \tag{SI2(5)}
\end{equation}%
where the $z$ integration was already performed, as in Eq.(SI2(2)). Changing 
$x\rightarrow -x$ in the second integral, one gets 
\begin{equation}
\frac{2}{R}\int_{x=R}^{\infty }\Sigma \left( x\right) \left( \frac{%
R^{2}-xR\cos \phi }{R^{2}-2Rx\cos \phi +x^{2}}-\frac{R^{2}+xR\cos \phi }{%
R^{2}+2Rx\cos \phi +x^{2}}\right) =2\pi Q\left( \phi \right) .  \tag{SI2(6)}
\end{equation}

Consider the right interface. Across the interface the current is
continuous. Both above and below the interface the current consists of the
solenoidal (transverse) and two irrotational (longitudinal) contributions
from the interface and the aperture surface. The upper is

\begin{equation}
J_{\phi }^{u}\left( x,0\right) =-\frac{\sigma _{T}\alpha }{2\pi x}+\sigma
_{L}2\pi \Sigma \left( x\right) +2\sigma _{L}\int_{\phi ^{\prime }=0}^{2\pi }%
\frac{R\sin \phi ^{\prime }Q\left( \phi ^{\prime }\right) }{%
x^{2}+R^{2}-2xR\cos \phi ^{\prime }},  \tag{SI2(7)}
\end{equation}%
while the lower is

\begin{equation}
J_{\phi }^{d}\left( x,0\right) =-\frac{\sigma \alpha }{2\pi x}-\sigma 2\pi
\Sigma \left( x\right) +2\sigma \int_{\phi ^{\prime }=0}^{2\pi }\frac{R\sin
\phi ^{\prime }Q\left( \phi ^{\prime }\right) }{x^{2}+R^{2}-2xR\cos \phi
^{\prime }}  \tag{SI2(8)}
\end{equation}%
Equating the two, one obtains:

\begin{equation}
\left( \sigma -\sigma _{T}\right) \frac{\alpha }{2\pi x}+2\pi \left( \sigma
+\sigma _{L}\right) \Sigma \left( x\right) -2\left( \sigma -\sigma
_{L}\right) \int_{\phi ^{\prime }=0}^{2\pi }\frac{R\sin \phi ^{\prime
}Q\left( \phi ^{\prime }\right) }{x^{2}+R^{2}-2xR\cos \phi ^{\prime }}=0 
\tag{SI2(9)}
\end{equation}

These two integral equations can be solved analytically. Rewriting
Eq.(SI2(8)) as 
\begin{equation}
\kappa _{T}\frac{\alpha }{2\pi x}+2\pi \Sigma \left( x\right) =2\kappa
_{L}R\int_{\phi =0}^{2\pi }\frac{\sin \phi Q\left( \phi \right) }{%
x^{2}+R^{2}-2xR\cos \phi },  \tag{SI2(10)}
\end{equation}%
with $\kappa _{T}\equiv \frac{\sigma -\sigma _{T}}{\sigma +\sigma _{L}}$ and 
$\kappa _{L}\equiv \frac{\sigma -\sigma _{L}}{\sigma +\sigma _{L}}$ and
substituting Eq.(SI3(7)) into the latter gives

\begin{equation}
\frac{\kappa _{T}\alpha }{2\pi x}+2\pi \Sigma \left( x\right) =\frac{2\kappa
_{L}}{\pi }\int_{\phi =0}^{2\pi }\frac{\sin \phi }{x^{2}+R^{2}-2xR\cos \phi }%
\int_{x^{\prime }=R}^{\infty }\Sigma \left( x^{\prime }\right) \left( \frac{%
R^{2}-x^{\prime }R\cos \phi }{R^{2}+x^{\prime 2}-2Rx^{\prime }\cos \phi }-%
\frac{R^{2}+x^{\prime }R\cos \phi }{R^{2}+x^{\prime 2}+2Rx^{\prime }\cos
\phi }\right) .  \tag{SI2(11)}
\end{equation}%
The integration over $\phi $ on the right hand side vanishes leading to

\begin{equation}
\Sigma \left( x\right) =-\frac{\kappa _{T}\alpha }{\left( 2\pi \right) ^{2}x}%
.  \tag{SI3(12)}
\end{equation}%
This combined with Eq.(SI2(7)) leads to the result Eq.(23).

\section{Appendix. The electric field at a circular aperture with an overall
neutral surface charge.}

A conducting cylinder with a circular aperture (radius $R)$ is charged
locally with surface charge density $Q\left( \phi \right) ,$ albeit overall
neutral: $\int_{\phi =0}^{2\pi }Q\left( \phi \right) =0$. We are interested
in the electric field just inside the conductor at $R+\Delta .$ The vector
from a point on the aperture surface to the observation point $\mathbf{r=}%
\left( R+\Delta \right) \left\{ \cos \chi ,\sin \chi \right\} $ is $\mathbf{s%
}=\mathbf{r-}R\mathbf{n}^{\prime },\mathbf{n}^{\prime }=\left\{ \cos \phi
,\sin \phi \right\} $. The (irrotational) field is given in SI2(2).

The normal to the aperture component just inside the metal is

\begin{equation}
\frac{\mathbf{r}}{R}\mathbf{\cdot E}^{irr}=2\int_{\phi =0}^{2\pi }\frac{%
\mathbf{r}^{2}\mathbf{-}R\mathbf{n}^{\prime }\mathbf{\cdot r}}{\mathbf{r}%
^{2}+R^{2}\mathbf{-2}L\mathbf{n}^{\prime }\mathbf{\cdot r}}Q\left( \phi
\right) =2\int_{\phi =0}^{2\pi }\frac{\left( L+\Delta \right) ^{2}\mathbf{-}%
L\left( L+\Delta \right) \cos \left( \phi -\chi \right) }{L^{2}+\left(
L+\Delta \right) ^{2}\mathbf{-2}L\left( L+\Delta \right) \cos \left( \phi
-\chi \right) }Q\left( \phi \right) .  \tag{SI2(13)}
\end{equation}%
After some algebraic manipulations one gets

\begin{equation}
\frac{\mathbf{r}}{R}\mathbf{\cdot E}^{irr}=2Q\left( \chi \right) \int_{\phi
=-\varepsilon }^{\varepsilon }\frac{\left( L+\Delta \right) ^{2}\mathbf{-}%
L\left( L+\Delta \right) \cos \left( \phi \right) }{L^{2}+\left( L+\Delta
\right) ^{2}\mathbf{-2}L\left( L+\Delta \right) \cos \left( \phi \right) }%
+2\int_{\phi =\varepsilon }^{2\pi -\varepsilon }\frac{\left( L+\Delta
\right) ^{2}\mathbf{-}L\left( L+\Delta \right) \cos \phi }{L^{2}+\left(
L+\Delta \right) ^{2}\mathbf{-2}L\left( L+\Delta \right) \cos \phi }Q\left(
\phi +\chi \right)  \tag{SI2(14)}
\end{equation}%
The integral in the first term is%
\begin{equation}
\int_{\phi =-\varepsilon }^{\varepsilon }\frac{\left( L+\Delta \right) ^{2}%
\mathbf{-}L\left( L+\Delta \right) \cos \left( \phi \right) }{L^{2}+\left(
L+\Delta \right) ^{2}\mathbf{-2}L\left( L+\Delta \right) \cos \left( \phi
\right) }=\pi +\varepsilon -2\arctan \frac{\Delta \cot \left( \varepsilon
/2\right) }{2L+\Delta },  \tag{SI2(15)}
\end{equation}%
and in the $\Delta \rightarrow 0$ limit $\pi +\varepsilon .$ In this limit
the second integral is simply $\int_{\phi =\varepsilon }^{2\pi -\varepsilon
}Q\left( \phi +\chi \right) .$

Taking now $\varepsilon \rightarrow 0$ and using neutrality one has the
result that the contribution from the circular surface is dominated by the
local charge:

\begin{equation}
\frac{\mathbf{r}}{R}\mathbf{\cdot E}^{irr}\mathbf{=}2\pi Q\left( \chi
\right) .  \tag{SI2(16)}
\end{equation}